\documentclass[floatfix,aip,jap,amsmath,amssymb,reprint]{revtex4-2}

\usepackage{amsfonts}
\usepackage{graphicx}
\usepackage{siunitx}
\usepackage{color}
\usepackage{physics} 
\usepackage{multirow} 
\usepackage[utf8]{inputenc}
\usepackage[T1]{fontenc}
\usepackage[colorlinks=true,citecolor=blue,urlcolor=red]{hyperref}

\begin{document}

\title{Force microscopy cantilevers locally heated in a fluid: temperature fields and effects on the dynamics}

\author{Basile Pottier}
\affiliation{Univ Lyon, Ens de Lyon, CNRS, Laboratoire de Physique, F-69342 Lyon, France}

\author{Ludovic Bellon}
\email{ludovic.bellon@ens-lyon.fr}
\affiliation{Univ Lyon, Ens de Lyon, CNRS, Laboratoire de Physique, F-69342 Lyon, France}

\date{\today}

\begin{abstract}
Atomic force microscopy cantilevers are often, intentionally or not, heated at their extremity. We describe a model to compute the resulting temperature field in the cantilever and in the surrounding fluid on a wide temperature range. In air and for common geometries, the heat fluxes in the cantilever and to the environment are of comparable magnitude. We then infer how the fluid-structure interaction is modified due to heating, and predict the induced changes in the dynamics of the system. In particular, we describe how the resonance frequencies of the cantilever shift with a temperature increase due to two competing processes: softening of the cantilever, and decrease of the fluid inertial effects. Our models are illustrated by experiments on a set of cantilevers spanning the relevant geometries to explore the relative importance of both effects.
\end{abstract}

\maketitle 

\section{Introduction}
In atomic force microscopy (AFM)\cite{Binnig-1986}, a sharp tip in interaction with the sample is used to characterise the surface topography or its physical properties\cite{Voigtlander-2019}. The tip-sample force transducer is a micro-mechanical device, commonly a cantilever, converting the force signal into a deflection that can be measured by high precision readouts\cite{Binnig-1986,Meyer-1988,Schonenberger-1989,Rugar-1989,Paolino-2013}. The understanding of the mechanical response of the transducer is thus crucial to fully interpret the final measurement. One distinctive feature of AFMs is the versatility of environments where they can operate, from vacuum to liquids, allowing many samples to be probed in their native state. In return, it complicates the characterisation of the force probe, since its response unavoidably implies its interaction with the environment. For cantilevers in fluids, for instance, the surroundings are paramount in the damping of the probe motion, but also have inertial effects visible on the mechanical resonances\cite{Butt-1993,Chen-1994,sader_frequency_1998}.

A complex situation is when the environment is not homogeneous: the fluid-structure coupling must in such a case take into account the variation of the fluid properties around the micro-mechanical sensor. In this article, we address the case of a non-uniform temperature for a cantilever in air. This scenario is naturally encountered in various situations, such as scanning thermal microscopy\cite{Gomes-2015}, where the AFM tip is heated to probe the local thermal conductivity or temperature of the sample, or as a consequence of photo-thermal excitation of a cantilever\cite{Marti-1992,Allegrini-1992,Ramos-2006,Kiracofe-2011,Bircher-2013} to drive it at its resonance frequency. Whereas the temperature field is a feature in these operation modes, in other cases it is an artifact of the detection methods. This is for example common for an optical readout since part of the light used to measure the deflection is absorbed by the cantilever and heats it. This effect is usually small but can get noticeable if strong light powers are needed (Raman measurements on the cantilever or sample\cite{McCarthy-2005,Lee-2006,milner_heating_2010,Chen-2011}, local infrared spectroscopy of the sample\cite{Lu-2014}, high thermal resistivity cantilevers\cite{aguilar_sandoval_resonance_2015}\ldots) In all cases, a precise knowledge of the temperature field in the cantilever and its surroundings is desirable, as well as understanding its effect on the mechanical response of the sensor.

In this article, we study the thermal and mechanical consequences of heating an AFM cantilever close to its free end in air. In previous works\cite{aguilar_sandoval_resonance_2015,Pottier-2021-JAP}, we tackled the problem in vacuum, which lifts two key difficulties: no heat flow to the environment, no hydrodynamic coupling. In air, the first point makes it harder to describe the temperature field in the cantilever and surroundings, as heat conduction/convection from a solid surface to a fluid is tricky to describe quantitatively. The second point has been solved by Sader and coworkers\cite{sader_frequency_1998,van_eysden_resonant_2006} for a uniform fluid, but the variation of density and viscosity around a heated cantilever adds an extra layer of complexity to the problem at stake. To address those difficulties, we present in this article a thorough study including a complete theoretical description and experimental results corroborating our main claims. We reach a precise understanding of the temperature field, a provide a quantitative picture of the fluids-structure effects on the mechanical response of the cantilever.

The article is organised as follows: in section \ref{section::thermalmodel}, we first present a model of the thermal problem and its resolution, leading to the knowledge of the temperature field everywhere. We successfully compare our model with a 3D simulation.  In section \ref{ModelFreqShift}, we describe the effect of a temperature field on the cantilever itself (in vacuum), and then how Sader's approach can be adapted to the varying properties of air heated in the vicinity of the cantilever. We end up with a model describing the inertial effects due to the fluid. In section \ref{section::exp.results}, we finally present an experiment to test our prediction, showing that measurements on a set of three cantilevers of different geometries in air and vacuum match nicely the proposed framework. The last section concludes this article with a discussion of the results and the perspectives opened by this work.

\section{Thermal problem: Cantilever heated by laser irradiation}\label{section::thermalmodel}

\subsection{In vacuum}

\begin{figure}[htb]
 \centering
 \includegraphics[scale=0.17]{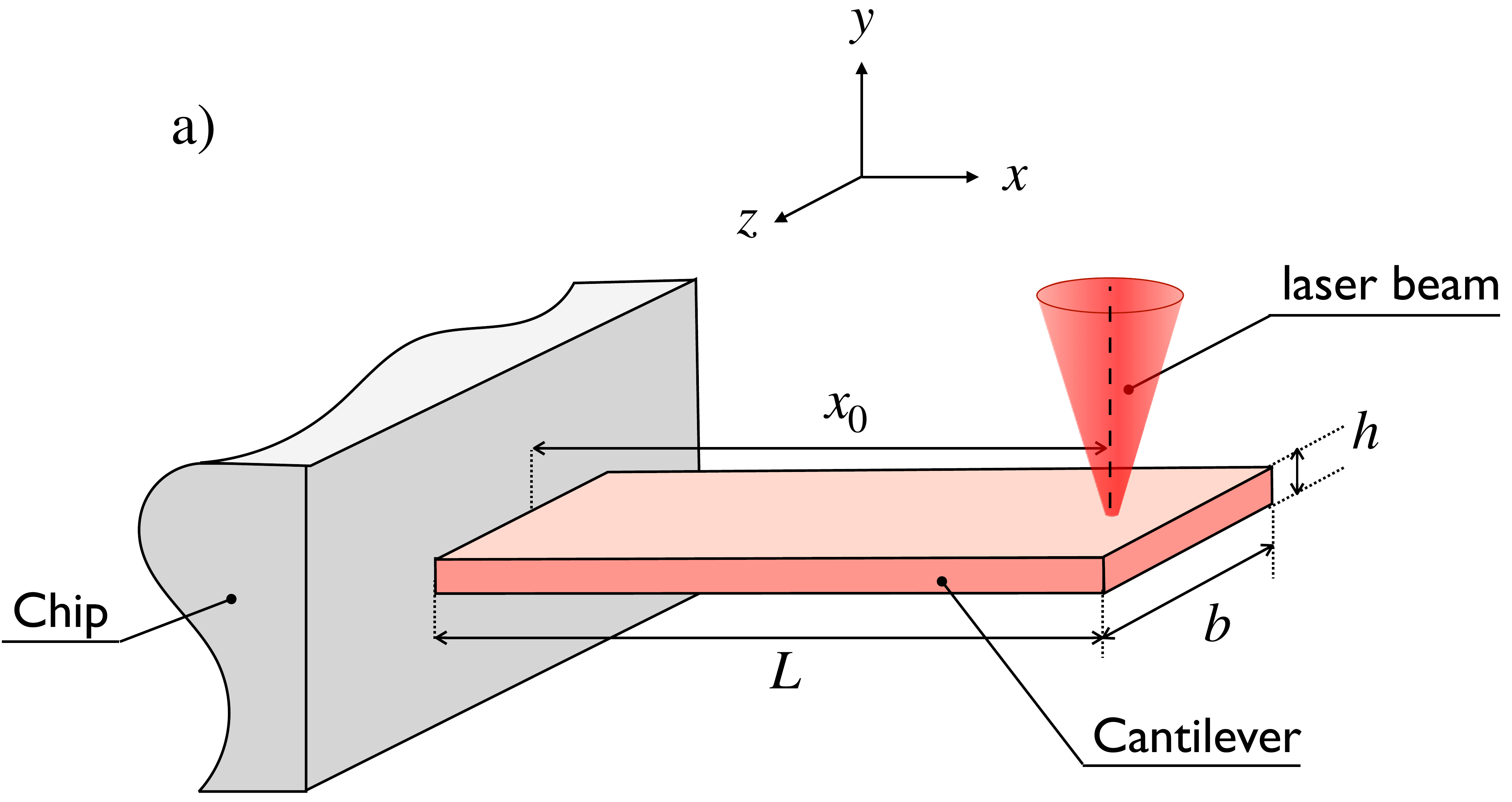}
 \vspace{3mm}
  \includegraphics[scale=0.17]{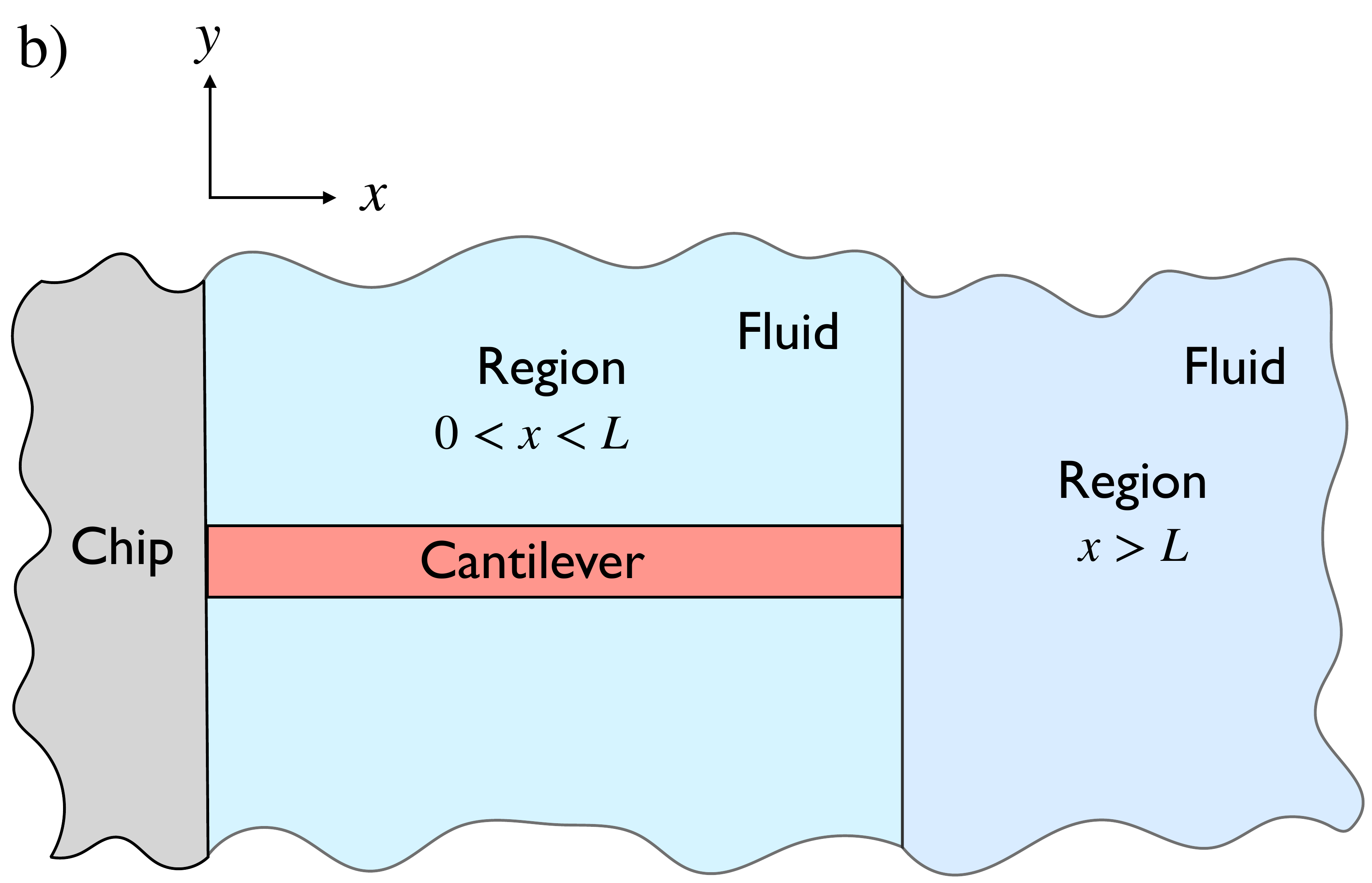}
  \caption{\label{FigSchemaCantPrism} a) Schematic representation of the rectangular cantilever with a clamped end to the chip at $x=0$ and a free end at $x=L$. b) Decomposition of the fluid volume surrounding the cantilever into two distinct regions.}
\end{figure}

Let us first consider the case when the cantilever heated by a laser beam is placed in vacuum. In this case, the only possible heat transfer mechanisms to dissipate the absorbed heat are thermal conduction through the cantilever and thermal radiation. For a rectangular cantilever whose cross section dimensions (width $b$, thickness $h$) are small compared to its length $L$ (figure~\ref{FigSchemaCantPrism}-a), the temperature can be assumed uniform across the cantilever cross section, the steady cantilever temperature profile $T(x)$ is thus solution of the one-dimensional equation
\begin{align}\label{HeatDiffVac}
    S\dv{}{x}\left( \lambda (T(x))\dv{T}{x} \right)+{q'}_\mathrm{laser} (x)-{q'}_\mathrm{rad} (x)=0,
\end{align}
where $S=bh$ is the cross section area, $\lambda$ is the cantilever thermal conductivity, ${q'}_\mathrm{laser}(x)$ and ${q'}_\mathrm{rad} (x)$ denote respectively the input and output lineic power (in $\si{W/m}$) owing to laser and cantilever radiation. For a laser beam whose beam size is small compared to the cantilever length, we can write ${q'}_\mathrm{laser}(x)=P_a\delta_D(x-x_0)$ where $P_a$ is the absorbed power, $\delta_D$ is Dirac's distribution and $x_0$ is the laser position along cantilever length. For a cantilever such that $b \gg h$, the lineic radiated power by the cantilever reads as ${q'}_\mathrm{rad} (x)=2\varepsilon\sigma b (T(x)^4-T_0^4) $ with $\varepsilon$ the material emissivity, $\sigma$ the Stefan-Boltzmann constant and $T_0$ the temperature of the environment. Note that for moderate temperature elevation, the radiative effects are small compared to thermal conduction\cite{Pottier-2021-JAP} and can be neglected.
 The temperature profile may be obtained by integrating twice Eq.~\eqref{HeatDiffVac} and imposing adequate boundary conditions. As discussed in Ref. \onlinecite{Pottier-2021-JAP}, the dissipated power inside the chip supporting the cantilever inevitably results in a rise of the cantilever temperature at the clamp $x=0$. This effect on the cantilever temperature can be characterized by the distance $l_\textrm{th}$ from the clamp where the cantilever temperature would extrapolate to $T_0$. The boundary condition at the clamped edge can thus be written as 
 \begin{align}\label{eq::CL_clamped}
	T(x=0)&= l_\textrm{th} \left. \dv{T}{x}\right\rvert_{x=0}+T_0.   
\end{align}
 In practice, $l_\textrm{th}/L$ is only about a few percent, the chip heating has thus a small effect on the behavior of the system but it has to be taken into account if one wants to accurately describe the temperature profile close to the clamp. At the opposite extremity, the free end for a cantilever in vacuum is thermally insulated and reads as
 \begin{align}\label{eq::CL_free}
	\left. \dv{T}{x}\right\rvert_{x=L}& =0.
\end{align}

\subsection{In a fluid}

When the cantilever is in contact with a gas or a liquid, heat will be transferred from the cantilever to its surrounding. Usually, the heat flux $q''$ transferred through a surface at temperature $T$ from a solid to the surrounding fluid at temperature $T_0$ is evaluated from Newton's law of cooling $q''=h_\mathrm{N}(T-T_0)$, where $h_\mathrm{N}$ is the effective heat transfer coefficient (in \si{W/m^2/K}). The coefficient $h_\mathrm{N}$ takes into account both heat transfer by conduction and convection through the fluid, it depends on the fluid properties and the specific geometry of the problem. For ideal configurations or geometries (as plane, cylinder or sphere) $h_\mathrm{N}$ can be calculated from formulas given in the literature \cite{incropera_fundamentals_2007}. For micron-scale devices, like AFM cantilevers, the buoyancy forces are not sufficiently large to overcome viscous forces\cite{lee_thermal_2007,kim_thermal_2009}. Thus, conduction dominates over convection for heat flow from the cantilever to the surrounding fluid. In that case, where natural convection can be neglected, the heat transfer coefficient scales as 
\begin{align}\label{eq::hscale}
    h_\mathrm{N}&\sim \frac{\lambda_\mathrm{fluid}}{l},
\end{align}
where $\lambda_\mathrm{fluid}$ is the thermal conductivity of the surrounding fluid and $l$ is the characteristic length of the temperature field. For a cantilever such that $b\ll L$, the fluid temperature will vary slowly along the length of the cantilever in comparison to transverse variations (in the plane $y-z$), the dominant length scale $l$ is thus the cantilever width $b$. Using Eq.~\eqref{eq::hscale} with $l=b$, we can roughly evaluate the effect of the heat transfer through the fluid relatively to the ones through the cantilever computing the dimensionless number $N=2\lambda_\mathrm{fluid}L^2/\lambda bh$. For a silicon cantilever with  $L=\SI{300}{\mu m}$, $b=\SI{30}{\mu m}$, $h=\SI{1}{\mu m}$ immersed in air, we get $N\approx 1$. Despite the low air conductivity relatively to silicon conductivity ($\lambda_\mathrm{Si}/\lambda_\mathrm{air}=\si{6\times 10^3}$), the very high aspect ratio of the cantilever geometry  ($L^2/bh=\si{3\times 10^3}$) makes the conduction effects in air be of the same order as those of conduction through the cantilever. It shows the importance to properly take into account the exchanged heat with air to accurately predict the cantilever temperature profile.

In the following, we present a model for the fluid temperature surrounding the cantilever that will allow deriving the corresponding local heat flux. Owing to the geometry of the problem, it makes sense to split the fluid surrounding the cantilever into two distinct sub-regions as illustrated in figure~\ref{FigSchemaCantPrism}-b: one region corresponds to the fluid between the chip ($x=0$) and the cantilever extremity ($x=L$), the second region corresponds to the remaining fluid $x>L$. In the following, we will treat separately the conduction problem in each region choosing appropriate boundary conditions. In a first step, we will address the problem in the particular case the cantilever is circular in cross section, allowing us derive analytical expressions for the heat exchanged. Then, we will give approximate expressions in our case of interest namely for a rectangular cross section. 

\subsubsection{Heat loss in the region \texorpdfstring{$x<L$}{x<L}}

For a cantilever having a circular cross section with a radius $R$ much smaller than its length $L$, the fluid temperature will vary slowly along the length of the beam in comparison to the transverse variations. It then follows that the temperature of the fluid locally surrounding the cantilever at any position $x$ along the cantilever length can be well approximated by that of an infinitely long cantilever of temperature $T(x)$, reducing the problem to two dimensions. The thermal flux $q''$ (in \si{W/m^2}) in the fluid is thus inversely proportional to the distance $r$ (with $r$ in cylindrical coordinates) and reads for a small cantilever element $dx$ as  
 \begin{align}\label{eqFourierX0L}
	q''=\frac{\dot{Q}_\mathrm{fluid}^{x<L}}{2\pi r dx}=-\lambda_\mathrm{fluid}  \frac{\partial T_\mathrm{fluid}^{x<L}}{\partial r},  
\end{align}
where $\dot{Q}_\mathrm{fluid}^{x<L}$ is the heat power (in \si{W}) transferred from the cantilever element to the surrounding fluid. Imposing the boundary conditions that (i) the fluid temperature equals the cantilever temperature at the contact ($r=R$), (ii) the fluid temperature at some distance $R_\infty$ is assumed to be equal to the reference temperature $T_0$, the integration of the Fourier law Eq.~\eqref{eqFourierX0L} gives the temperature of the fluid 
 \begin{align}\label{TempxsubL}
  T_\mathrm{fluid}^{x<L}(x,r)=\frac{\theta(x) }{\ln\left(\frac{R_\infty}{R}  \right)}\ln(\frac{R_\infty}{r})+T_0, \quad 0<x<L,
\end{align}
where $\theta(x)$ is the temperature elevation profile of the cantilever. Though we consider the 2D thermal problem perpendicular to the cantilever, a reasonable choice for $R_\infty$ is $R_\infty=R+x$, since $x$ is the distance of the cantilever element $dx$ from the chip acting as a thermostat at $T_0$. The power exchanged at position $x$ is thus  $\dot{Q}_\mathrm{fluid}^{x<L}=2 \pi dx\lambda_\mathrm{fluid}/\ln(1+x/R) \theta(x)$. The effect of heat conduction through the fluid in the region $0<x<L$ can thus be taken into account in the determination of the cantilever temperature by adding in Eq.~\eqref{HeatDiffVac} the local heat loss per unit of length term 
\begin{align}\label{Qfluidcirc}
	{q'}_\mathrm{fluid}^{ \, x<L}(x)= \frac{2 \pi}{ \ln(1+x/R)}\lambda_\mathrm{fluid} \theta(x).
\end{align}
Note that this expression correspond to Newton’s law of cooling with $h_\mathrm{N}= \lambda_\mathrm{fluid}/[R \ln(1+x/R)]$, close to the scaling expected from Eq.~\ref{eq::hscale}.
\subsubsection{Heat loss in the region \texorpdfstring{$x>L$}{x>L}}

In the previous paragraph, we only took into account the dissipation through the fluid located between the extremity of the cantilever and the chip ($0<x<L$). Here we will estimate the fluid temperature in half-space $x>L$ and give the associated heat flux. The problem can be reduced to a point source (the cantilever extremity) dissipating in the semi-infinite domain $x>L$, the thermal flux $q''$ is thus inversely proportional to $r^2$ (with $r$ in spherical coordinates) and reads as
\begin{align}\label{eqFourierXsupL}
    q''=\frac{\dot{Q}_\mathrm{fluid}^{x>L}}{2\pi r^2}=-\lambda_\mathrm{fluid}\frac{\partial T_\mathrm{fluid}^{x>L}}{\partial r},
\end{align} 
where $\dot{Q}_\mathrm{fluid}^{x>L}$ is the power dissipated to be determined. Imposing the boundary conditions that (i) at sufficiently far distances from the cantilever extremity the fluid is being cooled at the reference temperature ($T_\mathrm{fluid}(r\rightarrow \infty) =T_0$), (ii) the fluid temperature at $r=R$ equals the temperature of the cantilever at its extremity, the integration of Eq.~\eqref{eqFourierXsupL} gives 
\begin{align}\label{TempxsupL}
	T_\mathrm{fluid}^{x>L}(r)=\theta(L)\frac{R}{r}+T_0, \quad x>L,
\end{align}   
with $\theta(L)$ the temperature elevation of the cantilever at the cantilever extremity $x=L$. The power exchanged between the cantilever and the fluid in the region $x>L$ is thus  $\dot{Q}_\mathrm{fluid}^{x>L}=2\pi R \lambda_\mathrm{fluid}\theta(L)$. The effect of heat conduction through the fluid in the region $x>L$ can thus be taken into account in the determination of the cantilever temperature by imposing that the heat flux at the cantilever extremity is equal to
\begin{align}\label{jxsupL}
      {q''}_\mathrm{fluid}^{ \, x>L}&=\frac{\dot{Q}_\mathrm{fluid}^{x>L}}{S}=\frac{2\pi R}{S}\lambda_\mathrm{fluid}\theta(L),
\end{align}
where $S$ denotes the cross section area of the cantilever. Again, this expression correspond to Newton’s law of cooling with $h_\mathrm{N}= 2 \lambda_\mathrm{fluid}/R$, close to the scaling expected from Eq.~\ref{eq::hscale}.

\subsubsection{Temperature elevation of a cantilever in air}

Finally, the effect of heat loss through conduction within the fluid can be taken into account in the determination of the cantilever temperature $T(x)=\theta(x)+T_0$ by solving the modified heat equation 
\begin{align} \label{eqHeatfluid}
    S\dv{}{x}\left( \lambda (\theta)\dv{\theta}{x} \right)+{q'}_\mathrm{laser} (x)-{q'}_\mathrm{rad} (x)-{q'}_\mathrm{fluid}^{ \, x<L} (x)=0,
\end{align}
with the boundary conditions
\begin{subequations} 
\begin{align}
	\theta(x=0)&= l_\textrm{th} \left. \dv{\theta}{x}\right\rvert_{x=0}, \\ 
     \lambda(\theta(L)) \left. \dv{\theta}{x}\right\rvert_{x=L} &={q''}_\mathrm{fluid}^{ \, x>L},
\end{align}
\end{subequations}
where $ {q'}_\mathrm{fluid}^{ \, x<L}(x) $ and ${q''}_\mathrm{fluid}^{ \, x>L}$ are given in the case of a cantilever having a circular cross-section respectively by Eq.~\eqref{Qfluidcirc} and Eq.~\eqref{jxsupL}. It should be noted that the determination of the fluid temperature in each region that led up to these expressions implicitly assumed adiabatic the plane $x=L$, neglecting the temperature variation along $x$ in comparison to its transverse variations. This assumption and the resulting heat equation Eq.~\eqref{eqHeatfluid}  are thus valid in our limit where the cantilever transverse dimensions are small compared to its length.

For a cantilever having a rectangular cross section such that $b\ll h$, the width $b$ is the dominant length scale for the fluid temperature (instead of $R$ for a cylinder). The expressions for $ {q'}_\mathrm{fluid}^{ \, x<L}(x) $ and ${q''}_\mathrm{fluid}^{ \, x>L}$ should take a form similar to Eq.~\eqref{Qfluidcirc} and Eq.~\eqref{jxsupL} where the radius $R$ has been substituted by the half-width $b/2$, up to numerical factors of order one. We set those factors using 3D numerical simulations (presented in section \ref{sec:3Dsim}):
\begin{subequations} \label{eq::qcant}
\begin{align}
		{q'}_\mathrm{fluid,rect}^{ \, x<L}(x)	& \approx  \frac{2 \pi}{ \ln(1+4x/b)}\lambda_\mathrm{fluid} \theta(x),\\
		 {q''}_\mathrm{fluid,rect}^{ \, x>L} &\approx  \frac{b}{S}\lambda_\mathrm{fluid}\theta(L).
 \end{align}
\end{subequations}
The ratio of these two contributions to the thermal flux in air is of order $b/L$. For the typical geometries we consider, this ratio is around 0.1, so that the tip effect, though small, is usually noticeable and should be taken into account.

\begin{figure}[htb]
 \centering
  \includegraphics{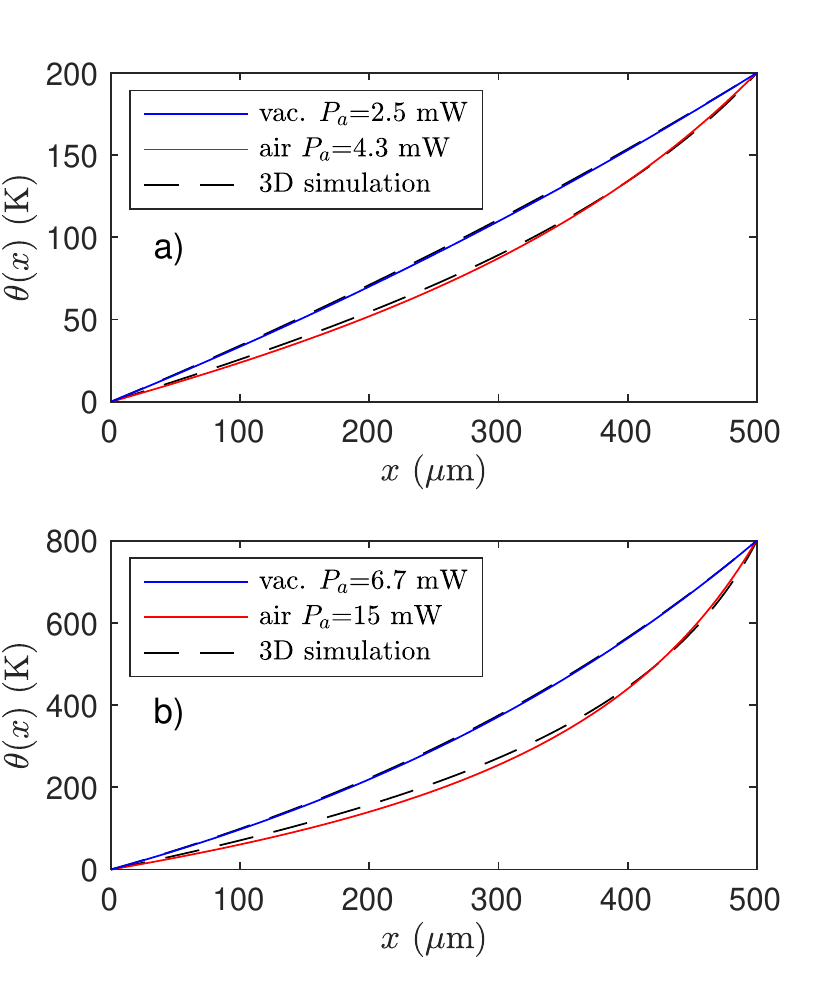}\vfill
  \includegraphics{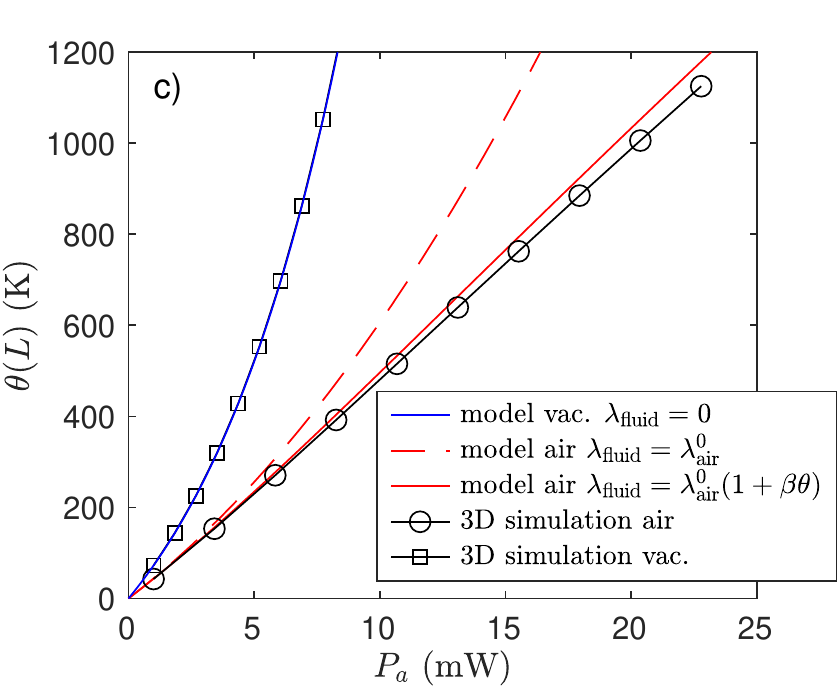}
  \caption{\label{fig::tempprofiles}Comparison of the temperature elevation profiles $\theta(x)$ computed for a silicon cantilever ($bh=\SI{50}{\mu m^2}$) placed in vacuum or in air for a given maximum temperature elevation $\theta(L)$ of 200~K (a) and 800~K (b). The heat exchanged with air enhances the non-linearity of the temperature profile. (c)~Temperature elevation at the tip of the cantilever $\theta(L)$ as a function of the absorbed power $P_a$ when the cantilever is placed in vacuum or air.}
\end{figure}
So far, the thermal conductivity for the fluid $\lambda_\mathrm{fluid}$ was implicitly assumed to be independent of the temperature. For a strong heating, this assumption may not be longer valid. At the first order, one can take into account the temperature dependence of the fluid conductivity with $\lambda_\mathrm{fluid}( \theta)=\lambda_\mathrm{fluid}^0\left(1+\beta \theta \right)$, with $\beta=\SI{1/410}{K^{-1}}$ for air. Recasting this expression of $\lambda_\mathrm{fluid}$ in the computations leading to Eqs.~\eqref{eq::qcant}, the resulting expressions for ${q'}_\mathrm{fluid}$ and ${q''}_\mathrm{fluid}$ are unchanged, simply substituting the constant thermal conductivity by $\lambda_\mathrm{fluid}^0 \left(1+\beta \theta(x)/2 \right)$.

In the following, we compare the temperature elevation profiles $\theta(x)$ of a silicon cantilever placed in vacuum or in air predicted by our model solving Eq.~\eqref{eqHeatfluid} when the laser is focused at its free end ($x_0=L$). We take into account the temperature dependence of the thermal conductivities $\lambda$ of silicon\cite{glassbrenner_thermal_1964} and $\lambda_\mathrm{fluid}$ of air conductivity at first order. The profiles are computed neglecting both the effect of radiation (${q'}_\mathrm{rad}=0$) and chip heating ($l_\mathrm{th}=0$). In figure~\ref{fig::tempprofiles}, we display the profiles $\theta(x)$ computed for a maximum temperature elevation of $\SI{200}{K}$ and $\SI{800}{K}$. In vacuum, the non-linearity of the profile comes from the temperature dependence of the silicon conductivity. Due to the heat exchanged with its surrounding, the profiles in air deviate from the ones in vacuum, with a more pronounced non-linearity. The absorbed power $P_a$ needed to get a temperature rise at the extremity of $\SI{200}{K}$ and $\SI{800}{K}$ is respectively 1.7 and 2.3 times larger in air than in vacuum. In figure~\ref{fig::tempprofiles}-c, we display the maximum temperature elevation $\theta(L)$ in air and in vacuum as a function of the absorbed power $P_a$. For comparison we display the temperature rise assuming constant the thermal conductivity of air.

\subsection{Validation with 3D simulation} \label{sec:3Dsim}

\begin{figure}[htb]
 \centering
 \includegraphics[width=\columnwidth]{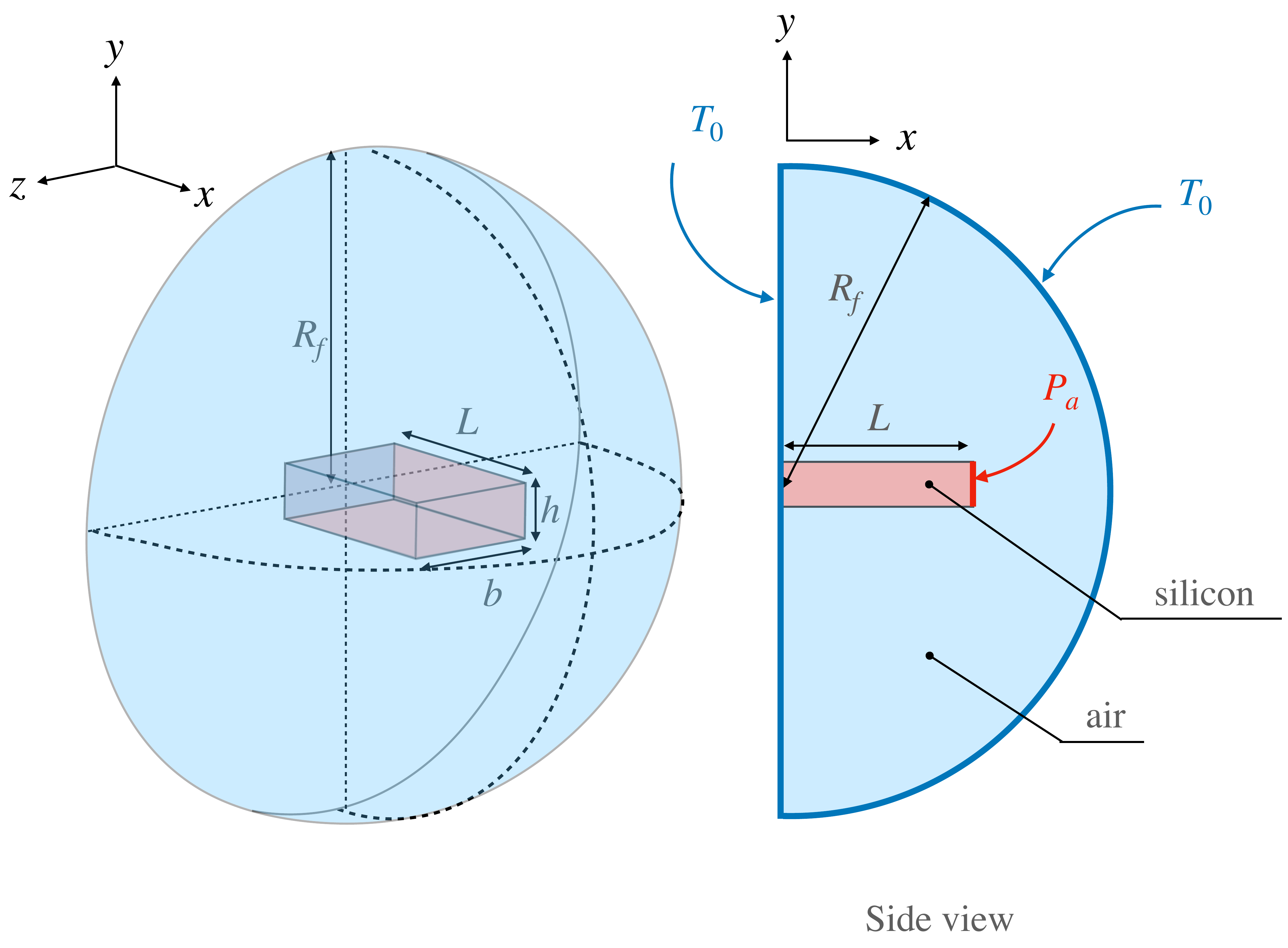}
  \caption{\label{fig::3DcantSimu} Schematic of the 3D thermal simulation of the cantilever and its surrounding, not to scale. The silicon cantilever is a prismatic volume ($h \times b \times L$) while the surrounding air is within the hemisphere of radius $R_f$. All exterior surfaces of the simulated volume are maintained at a constant temperature $T_0$. The power $P_a$ is introduced within the cantilever at its free end $x=L$.}
\end{figure}

To check the validity of our thermal model, we perform a 3D numerical simulation of the conduction problem of a silicon cantilever surrounded by air. In the simulation presented, the only thermal transfer process is conduction. Figure~\ref{fig::3DcantSimu} shows the geometry of the simulated problem. The cantilever consists of a rectangular parallelepiped ($L \times b \times h $) and its surrounding environment consists of the remaining volume within the hemisphere of radius $R_f$ centered at the clamped end. The heat equation solved in the two volumes is performed using the thermal conductivity data of silicon for the cantilever and air for the surrounding, taking into account their temperature dependence. We impose the boundary condition of an isothermally surface (at the reference temperature $T_0$) at all exterior surfaces of the simulated volume. The absorbed power $P_a$ is poured within the cantilever at its free end $x=L$. The simulation is performed using COMSOL Multiphysics. 

We present the result for a cantilever such that $h=\SI{1}{\mu m}$, $b=\SI{50}{\mu m}$, $L=\SI{500}{\mu m}$ with $R_f=5L$. We verified that the obtained cantilever temperature profile is not affected by simulating a larger air domain (such that $R_f>5L$). In figure~\ref{fig::tempprofiles}, we display the cantilever temperature profiles $\theta(x)$ and the variation of the temperature at its free end $\theta(L)$ as a function of the absorbed power $P_a$ both taking into account the heat transfer with air or not. The temperatures obtained from the simulation are in excellent agreement with the predictions made by our model presented above. We find such agreement between 3D simulation and our model using different cantilever dimensions $b$, $h$ and $L$ as long as $b \gg L$ and  $b \gg h$. 

In conclusion, our model can be used to quantitatively determine the temperature distribution along the cantilever when it is placed either in vacuum or in air.

\section{Mechanical problem: frequency response of a cantilever submitted to a temperature profile}\label{ModelFreqShift}

We now present a general theoretical model to determine the resonance frequency and quality factor of a cantilever beam immersed in a viscous fluid submitted to a one-dimensional temperature profile. For a rectangular cantilever such that its length $L$ greatly exceeds its transverse dimensions (see figure~\ref{FigSchemaCantPrism}-a), the dynamic motion of the cantilever undergoing small flexural deformation can be described in the Euler-Bernoulli framework. The equation governing the deflection $w(x,t)$ reads \cite{timoshenko_vibration_1937}
\begin{align}\label{EqEBtemp}
	  \mu_\mathrm{c} \frac{\partial^2 w}{\partial t^2}+ \frac{\partial^2}{\partial x^2}    \left[ EI   \frac{\partial^2w}{\partial x^2}   \right]  = f_\mathrm{fluid}(x,t)+f_\mathrm{ext}(x,t),
\end{align} 
where $\mu_c=\rho_c bh$ is the mass per unit length of the cantilever of density $\rho_c$, $E$ is the Young's modulus, $I$ is the moment of inertia of the beam, $f_\mathrm{fluid}$ is the hydrodynamic force per unit length (in \si{N/m}) due to the surrounding fluid (if any) acting on the cantilever, and $f_\mathrm{ext}$ is the external force per unit length. All quantities $\mu_c$, $E$ and $I$ in Eq.~\eqref{EqEBtemp} may depend on the position $x$ along cantilever length. The boundary conditions are the usual clamped-free conditions: 
\begin{subequations}
\begin{alignat}{2}
    w(x=0,t) &=0, \quad \left.\frac{\partial w(x,t)}{\partial x}\right|_{x=0} &&=0, \\
    \left.\frac{\partial^2 w(x,t)}{\partial x^2}\right|_{x=L} &=0, \quad \left.\frac{\partial^3 w(x,t)}{\partial x^3}\right|_{x=L} &&=0.
\end{alignat}
\end{subequations}

To solve Eq.~\eqref{EqEBtemp}, it is convenient to expand the deflection $w(x,t)$ in terms of the eigenmodes $\phi_n$ of the bare cantilever (without fluid and external forces) having uniform properties. We therefore write in Fourier space
\begin{align}
    \hat{w}(x,\omega) &= \sum_{n=1}^{\infty} W_n(\omega)\phi_n(X), \quad X=x/L,
\end{align}
where $n$ is the mode order. Similar  expression holds for the external force $f_\mathrm{ext}(x,\omega)$. The eigenmodes $\phi_n(X)$ satisfy the following conditions
\begin{subequations}
\label{EqNormalMode}
\begin{align}
 \phi''''_n(X)-\alpha_n^4 \phi_n(X)=0, \\
\phi_n(0)=\phi'_n(0)=\phi''_n(1)=\phi'''_n(1)=0,
\end{align}
\end{subequations}
where $\alpha_n$ are the successive positive roots of
\begin{align}\label{alphan}
    1+\cos \alpha_n \cosh \alpha_n=0.
\end{align}

For small deflection, the hydrodynamic force per unit length $f_\mathrm{fluid}$ is proportional to the local displacement of the cantilever $w(x,t)$ \cite{sader_frequency_1998}. We therefore write in Fourier space
\begin{align}\label{EqForceHydro}
    \hat{f}_\mathrm{fluid}(x,\omega)&= \mu \omega^2 \Gamma (\omega)   \hat{w}(x,\omega),
\end{align}
where $\mu=\pi \rho b^2 /4$ is the mass per unit length of a cylinder of diameter $b$ of fluid with density $\rho$. $\Gamma$ is the dimensionless complex hydrodynamic function. The real part $\Gamma_r$ accounts for the inertial effects, while the imaginary part $\Gamma_i$ accounts for the viscous damping. Substituting Eq.~\eqref{EqForceHydro} into Eq.~\eqref{EqEBtemp} and rearranging we obtain in Fourier space
\begin{align}\label{EqEulerBerfeq}
	 \frac{ \dd^2 }{\dd x^2}    \left[ EI   \frac{\dd^2 \hat{w}}{\dd x ^2}   \right] -  \left(\mu_c+\mu\Gamma(\omega) \right) \omega^2  \hat{w} = \hat{f}_\mathrm{ext}(x,\omega).
\end{align}

\begin{figure}[htb]
 \centering
   \includegraphics{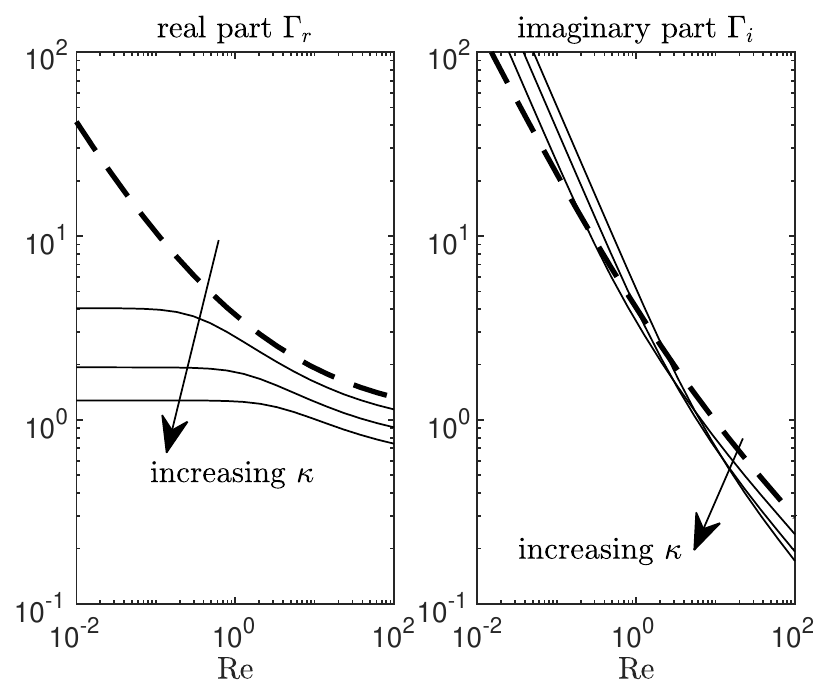}
  \caption{\label{FigGamma} Hydrodynamic function $\Gamma=\Gamma_r+i\Gamma_r$ for a rectangular cross section ($b\gg h$) as a function of the Reynolds number $\mathrm{Re}=\rho \omega b^2/4 \eta$ for varying wave number $\kappa=\alpha_n b/L=0,1,2,3$ computed from the analytical solution of Ref.~\onlinecite{van_eysden_small_2006}. The dashed line corresponds to the two-dimensional rigid cantilever case ($\kappa=0$).}
\end{figure}

For relatively low mode number $n$, the hydrodynamic flow can be assumed two dimensional in nature\cite{van_eysden_resonant_2006}. The hydrodynamic function $\Gamma$ is then obtained from the solution of the linearized Navier-Stokes equations for a rigid beam with identical cross section to that of the cantilever beam undergoing transverse oscillatory motion \cite{sader_frequency_1998}. For a rectangular cross section such that $b \gg h$, $\Gamma$ depends on the radial frequency $\omega$ through the Reynolds number $\mathrm{Re}=\rho \omega b^2/4 \eta$ where $\eta$ is the fluid viscosity and the cantilever width $b$ corresponds to the dominant length scale in the hydrodynamic flow \cite{tuck_calculation_1969} (figure~\ref{FigGamma}). For relatively high mode number, the flow in the fluid surrounding the cantilever can no longer be assumed two-dimensional. Sader and coworkers showed\cite{van_eysden_resonant_2006} that the hydrodynamic function thus depends on the angular frequency $\omega$ and the mode number $n$ through $\mathrm{Re}$ and the normalized mode number $\kappa = \alpha_n b/L$ where $\alpha_n$ is defined in Eq.~\ref{alphan} (see figure~\ref{FigGamma}). 

We now consider the effect of the temperature profile $T(x)=T_0+\theta(x)$ on the physical properties of the cantilever and the surrounding fluid (if not in vacuum). In the general case, all parameters $E$, $I$, $\mu_c$, $\mu$ and $\Gamma(\omega)$ in Euler-Bernoulli equation~Eq.~\eqref{EqEulerBerfeq} depend on the variable $x$ through their temperature dependency and may be expressed in the following form 
\begin{align}
    g(x)&=g_0+\Delta g(x),
\end{align}
where $g_0$ corresponds to the parameter value at the reference temperature $T_0$ and $\Delta g(x)$ corresponds the parameter variation at the position $x$ induced by the temperature profile $\theta(x)$. When the variations are relatively small, i.e. $\Delta g(x) \ll g_0$,
the normal modes are unchanged at the first order\cite{aguilar_sandoval_resonance_2015,Pottier-2021-JAP}. In this limit, the evolution of the amplitude of each mode can be obtained by projecting Eq.~\eqref{EqEulerBerfeq} on the known normal mode basis $\phi_n(X)$:
\begin{align}\label{TransferFunct}
     \left[  - 	 m_\mathrm{eff}^n \omega^2 + k_\mathrm{eff}^n + i\gamma_\mathrm{eff}^n \omega \right] W_n(\omega) = L F_n^\mathrm{ext}(\omega),
\end{align} 
with
\begin{subequations}\label{Eq::param_mkg}
\begin{align}
   m_\mathrm{eff}^n&=L\int_0^1  \left(\mu_c+ \mu\Gamma_r(\omega) \right) \phi_n(X)^2  \dd X,\label{Eq::param_m}\\
   k_\mathrm{eff}^n&=\frac{1}{L^3}\int_0^1 EI  \left. \phi''_n(X) \right.^2  \dd X, \\
   \gamma_\mathrm{eff}^n&=\omega L \int_0^1 \mu \Gamma_i(\omega) \phi_n(X)^2  \dd X.
\end{align} 
\end{subequations}

For the large quality factors typical in air and vacuum, the modes can be considered uncoupled to leading order. We therefore have a collection of independent quasi-harmonic oscillators of effective mass $m_\mathrm{eff}^n$, stiffness $k_\mathrm{eff}^n$ and damping coefficient $\gamma_\mathrm{eff}^n$. From Eqs.~\eqref{Eq::param_mkg}, both dissipative and inertial effects are weighted by the square amplitude of the normal mode, while the elastic effect is weighted by the square of the local curvature. We immediately identify from Eq.~\eqref{TransferFunct}, the eigenfrequency associated with each mode as 
\begin{align}\label{FreqResGener}
	\omega_n^2 & =\frac{k_\mathrm{eff}}{m_\mathrm{eff}} = \frac{ \displaystyle\int_0^1 E I \left. \phi_n''(X) \right. ^2 \dd X } { L^4 \displaystyle\int_0^1 \left( \mu_c+\mu  \Gamma_r(\omega_n) \right)  \left. \phi_n(X)   \right.^2 \dd X }, 
\end{align}
and the quality factor as $Q_n=m_\mathrm{eff}\omega / \gamma_\mathrm{eff}$. 

In the particular case both the cantilever and fluid properties are uniform, the resonance frequency of the cantilever immersed the fluid $\omega_n^\mathrm{fluid}$ given by Eq.~\eqref{FreqResGener} becomes 
\begin{align}\label{eq::FluidUnif}
  \omega_n^\mathrm{fluid}  &=\frac{\omega_n^\mathrm{vac}}{\sqrt{1+\zeta}},
\end{align} 
where
\begin{align}\label{EqVacUnif}
    \omega_n^\mathrm{vac}&=\frac{\alpha_n^2}{L^2}\sqrt{\frac{EI}{ \mu_c}},
\end{align} 
is the resonance frequency in vacuum, and
\begin{align}\label{eq::zeta}
\zeta=\frac{\pi \rho b}{4 \rho_c h}\Gamma_r(\omega_n^\mathrm{fluid})
\end{align} 
is a dimensionless number representing the inertial effects of the fluid on the resonance frequency. This effect is more pronounced on larger cantilevers (i.e. large aspect ratio $b/h$) since the volume of fluid implied is higher. For typical cantilever geometries ($10<b/h<100$) the relative frequency shift from vacuum to air is in the few percent range. The quality factor becomes
\begin{align}\label{eq::QualityFac}
    Q_n^\mathrm{fluid }=\frac{\frac{4 \rho_c h}{\pi \rho b}+\Gamma_r(\omega_n^\mathrm{fluid})}{\Gamma_i(\omega_n^\mathrm{fluid})}.
\end{align} 

\begin{figure}[htb]
 \centering
   \includegraphics{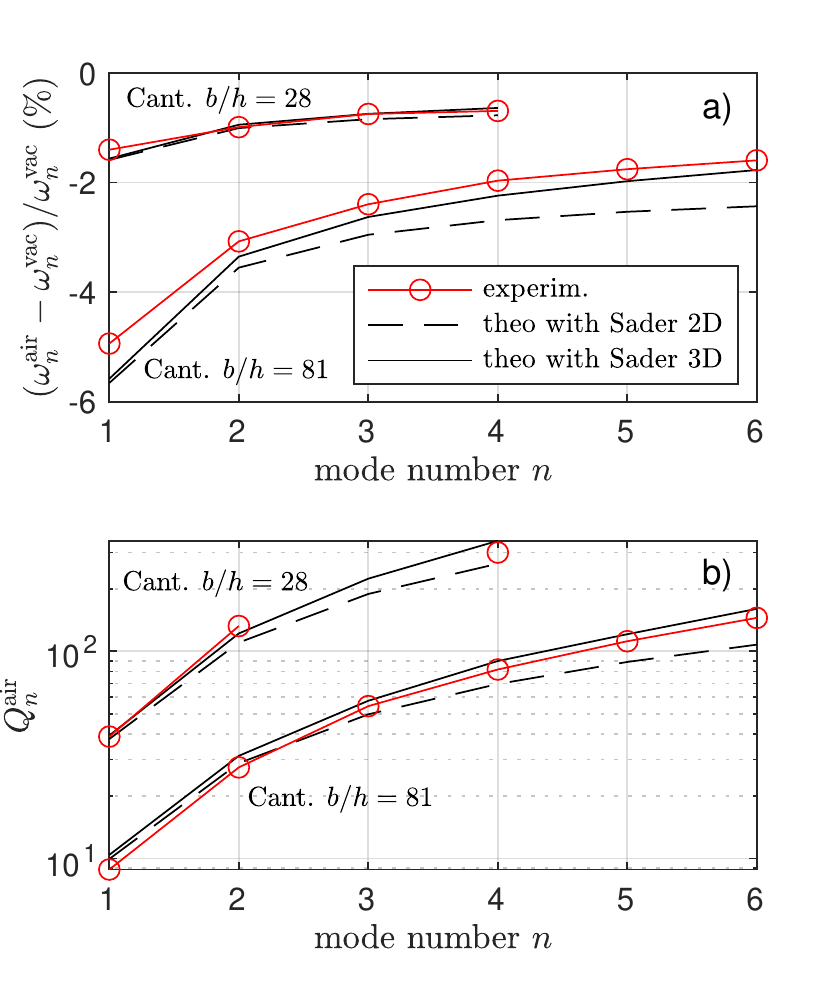}
  \caption{\label{CompRoomTemp} Relative frequency shift from vacuum to air $(\omega_n^\mathrm{air}-\omega_n^\mathrm{vac})/\omega_n^\mathrm{vac}$ (a) and quality factor\cite{NoteMissingQ} $Q_n^\mathrm{air}$ (b) in air measured at room temperature from a thermal noise spectrum for two silicon cantilevers having respectively the aspect ratio $b/h=28$ and $b/h=81$. Both the shift and the quality factor are well described with no adjustable parameters by Eq.~\eqref{eq::FluidUnif} and Eq.~\eqref{eq::QualityFac} with the hydrodynamic function $\Gamma$ computed from Ref.~\onlinecite{van_eysden_resonant_2006}. For high mode number, the air flow around the cantilever can no longer be assumed two dimensional.}
\end{figure}

For illustration, we display in figure~\ref{CompRoomTemp} this shift and the quality factor in air measured at room temperature with two cantilevers having different aspect ratios $b/h$. As expected, because of the frequency dependence of the hydrodynamic load described through $\Gamma_r$ (displayed in figure~\ref{FigGamma}), the shift slightly decreases with the considered mode number $n$. For high mode numbers, the fluid can be considered as inviscid, $\Gamma_r(\mathrm{Re}\rightarrow \infty)=1$, the frequency shift thus becomes independent of the mode number. We compare in figure~\ref{CompRoomTemp} the measured relative shift and the quality factor to the theoretical values given respectively by Eq.~\eqref{eq::FluidUnif} and Eq.~\eqref{eq::QualityFac} where $\Gamma$ is computed either considering the flow two dimensional or three dimensional\cite{van_eysden_resonant_2006}). As expected, for high mode number, the flow can no longer be assumed two dimensional, both the frequency shift and the quality factor slightly deviate from the 2D model and are better described by the 3D model.

In the following, we determine the sensitivity of the frequency resonance as the cantilever is submitted to a one-dimensional temperature profile $T_0+\theta(x)$.

\subsection{Sensitivity to a uniform temperature \texorpdfstring{$\theta(x)=\Delta T$}{} }

To forge our intuition, we first consider a small uniform temperature rise $\theta(x)=\Delta T$. When placed in vacuum, the cantilever temperature rise will induce a shit of the resonance frequency owing to the material softening (decrease of $E$) and an increase of the cantilever dimensions due to thermal dilatation. Differentiating Eq.~\eqref{EqVacUnif} with respect to the temperature around $T_0$, the relative frequency shift in vacuum is
\begin{align}\label{EqShiftVacUnif}
	\Delta \Omega_n^\mathrm{vac} \equiv\frac{\omega^\mathrm{vac}_n-\omega^{\mathrm{vac},0}_n}{\omega^{\mathrm{vac},0}_n}=\frac{1}{2}\left( a_E+a_l \right) \Delta T,
\end{align}
where the coefficient $a_E$ and $a_l$ denotes respectively the temperature coefficient of the Young’s modulus and the coefficient of linear thermal expansion. Since  $a_E=\SI{-64e-6}{K^{-1}}$ and  $a_l=\SI{2.5e-6}{K^{-1}}$ for silicon\cite{masolin_thermo-mechanical_2013,gysin_temperature_2004,okada_precise_1984,okada_precise_1984,watanabe_linear_2004}, the softening effect is dominant: a rise of temperature tends to decrease the resonant frequency.

When placed in a fluid, the temperature rise of the cantilever will heat the surrounding fluid, modifying both its viscosity and its density thus altering the inertial effect of the fluid on the resonance frequency. In air, the added mass effect is small since $\zeta  \ll 1$. To simplify, let us consider the inertial regime $\mathrm{Re}\gg 1$, i.e., $\Gamma_r \approx 1$. From Eq.~\eqref{eq::FluidUnif}, the relative frequency shift is then at first order
\begin{align}\label{ShiftFluidUnif}
	\Delta \Omega_n^\mathrm{fluid} &=\frac{1}{2} \big[ a_E +a_l -\zeta (a_\rho+3a_l) \big] \Delta T,
\end{align}
where $a_\rho$ denotes the volumetric coefficient of thermal expansion of the fluid. For air at room temperature, $a_\rho=\SI{-3.4e-3}{K^{-1}}\gg a_l$, so that for a silicon cantilever with $b/h=30$, we have $\zeta a_\rho /a_E=0.64$. The softening of the cantilever and the dilatation of the surrounding air are expected to have opposite effects of similar magnitude on the resonance frequency. We emphasize here that even if the presence of air changes only the resonant frequency by a few percent, it can dramatically change its sensitivity to a temperature change.

\subsection{Sensitivity to an arbitrary temperature profile}

In the general case where the temperature rise is not uniform, the Young's modulus $E$ as well as the cross-section dimensions, thus its second moment of inertial $I$ and lineic density $\mu_c$, are a function of $x$ through the temperature profile. Substituting the $x$ dependent parameters into Eq.~\eqref{FreqResGener}, the relative frequency shift in vacuum due to the temperature profile $\theta(x)=\Theta(x)\Delta T$ reads as  
\begin{align}\label{eqShiftVac}
\begin{split}
	\Delta \Omega^\mathrm{vac}_n & =   g_n(\Delta T)\\
	&= \frac{1}{2}\displaystyle\int_0^1 \left[  (A_E(\theta(X))+4A_l(\theta(X)))\frac{\left. \phi''_n (X)\right.^2}{\alpha_n^2}\right. \\ &\left. + A_l(\theta(X)) \left. \phi_n (X)\right.^2 - 4A_l(\theta(X))  \right] \dd X, 
\end{split}
\end{align}
with 
\begin{subequations}
\begin{align}
	A_E(\theta) &= \int_{T_0}^{T_0+\theta}  a_E(T)  \dd T =  \frac{E( T_0+\theta)}{E_0} -1,\\
	A_l(\theta) &= \int_{T_0}^{T_0+\theta}a_l(T) \dd T.  
\end{align}
\end{subequations}
The usefulness of Eq.~\eqref{eqShiftVac} has already been demonstrated and quantitatively compared to experiments in a previous work\cite{Pottier-2021-JAP}: knowing the mechanical properties ($a_E$, $a_l$) and the temperature distribution along the cantilever $\Theta(x)$, the function $g_n(\Delta T)$ in Eq.~\eqref{eqShiftVac} can be computed numerically for any mode number $n$ to predict the associated relative frequency shift $\Delta \Omega_n^\mathrm{vac}$. Conversely, it is possible to deduce the temperature increase $\Delta T$ from the relative frequency shift using $\Delta T=g_n^{-1}(\Delta \Omega_n^\mathrm{vac})$ where the $g_n^{-1}$  are the inverse functions of $g_n$.

In air, an additional term should be summed to Eq.~\eqref{eqShiftVac}:
\begin{align}\label{eqShiftAir}
    \Delta \Omega^\mathrm{fluid}_n & = \Delta \Omega^\mathrm{vac}_n-   \frac{1}{2}\frac{\zeta_0}{1+\zeta_0} \displaystyle\int_0^1   A_\zeta(\theta(X) ) \left. \phi_n (X)\right.^2  \dd X, 
\end{align}
where $\zeta_0$ is the inertial parameter due to the fluid at the reference temperature $T_0$, and $A_\zeta=(\zeta-\zeta_0)/\zeta_0$ its relative change due cantilever temperature increase $\theta(X)$. Note that this $\theta(X)$ dependency is complex: it includes the modifications of the fluid density $\rho$ by the temperature in the fluid, but also of the Reynolds number $\mathrm{Re}$ through $\rho$ and $\eta$, all depending on the distance to the cantilever in the fluid. This addition term is an inertial effect: the changes of the fluid properties influence the resonance frequency through modifications of $m_\mathrm{eff}^n$. Heating has thus potentially a greater influence on the first mode: the larger changes in the fluid properties occur where the cantilever is the hotter, hence where the weighting by the square amplitude of mode one is the largest in Eq.~\eqref{Eq::param_m}.

\section{Experimental results} \label{section::exp.results}

\subsection{Experimental setup}

\begin{figure}[tb]
 \centering
  \includegraphics[width=\columnwidth]{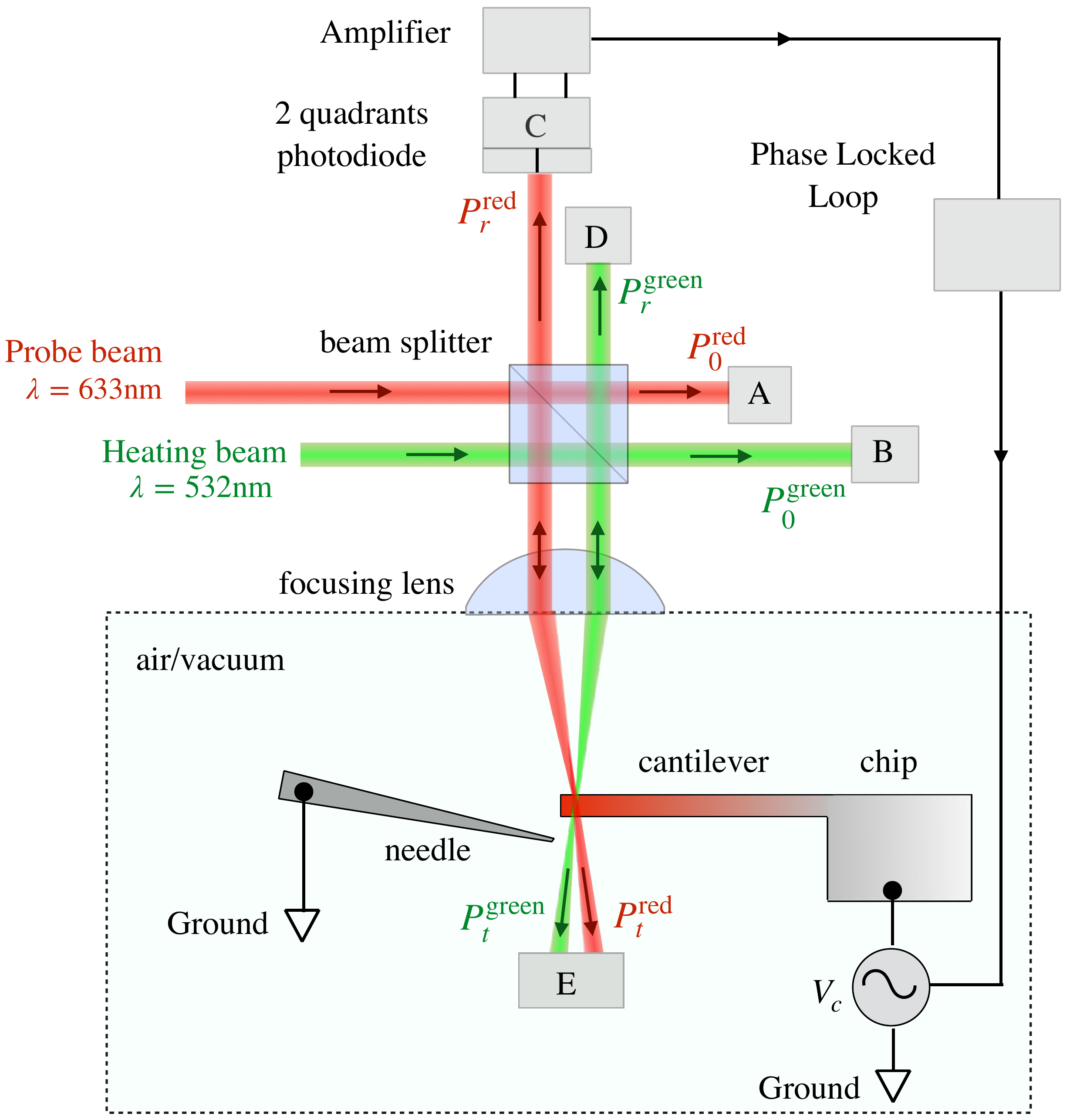}
  \caption{\label{Fig::SchemaExperiment} Experimental setup to heat the cantilever and track its resonance frequencies. The cantilever either in vacuum or air is illuminated by two laser beams focused at its extremity (focal length $\SI{30}{mm}$, spot size $5$ to $\SI{7}{\mu m}$). By varying the incident power of the 532~nm laser beam, the cantilever can be heated from room temperature up to several hundred degrees. The reflected beam of the 633~nm laser is sent in a two-quadrant photodiode allowing to detect cantilever deflection and to track its resonance frequencies. The cantilever is electrostatically actuated applying a voltage controlled by the phase locked loop (Nanonis OC4). Both laser beams are actually perpendicular to the cantilever and superposed, they have been separated in this sketch for illustration purposes. Separation before the photodiodes are performed thanks to dichroic filters.}
\end{figure}

To study experimentally the influence of air on the resonance frequency, we present in this section measurements on a heated silicon cantilever either placed in vacuum or air at ambient pressure (figure~\ref{Fig::SchemaExperiment}). We drive the resonances applying an electrostatic force, and read the corresponding deflection using an optical beam deflection scheme, with a low intensity He-Ne laser ($<\SI{1}{mW}$). The resonance frequency is tracked with a phase locked loop (PLL). Heating is induced by partial absorption of a second laser beam (0 to $\SI{50}{mW}$ at $\SI{532}{nm}$), focused close to the free end of the cantilever. After proper calibration, a set of 5 photodiodes (A to E, including the 2-quadrants photodiode C used to measure deflection) allows the measurement of the incident light power of each beams $P_0^\mathrm{green}$, $P_0^\mathrm{red}$, of the reflected one $P_r^\mathrm{green}$ and $P_r^\mathrm{red}$, and of the light transmitted through the cantilever $P_t^\mathrm{green}$ and $P_t^\mathrm{red}$. Indeed, a few micrometers thick silicon cantilever is semi-transparent for those wavelengths, its absorption coefficient results from the interferences within the cantilever and is expected to vary during the experiment through the temperature\cite{Pottier-2021-JAP,Pottier-2021-SciPost}. Since light diffusion is negligible in this experiment, those measurements allow deducing the total absorbed power $P_a$ by the cantilever during the whole measurement. The cantilever is placed in a chamber filled with air either at atmospheric pressure or at \SI{1e-4}{mBar}. At the latter pressure level, the contribution of convective heat transfer is negligible compared to thermal conduction\cite{lee_thermal_2007}. 

As the width to thickness $b/h$ is the relevant aspect ratio to study the influence of the surrounding fluid on the resonance frequency, we measure three cantilevers (see~table \ref{table::GeomCant}): Cant-A (BudgetSensors AIO-CM), Cant-B (Mikromasch CSC38-B) and Cant-C (Micromotive octo-500) having a width to thickness ratio $b/h$ respectively of 11, 28, and 83. All are uncoated tipless atomic force microscope silicon cantilevers. For each tracked resonance, the incident power of the 532~nm beam is continuously increased up to a maximal value then symmetrically decreased. The duration of one measurement is approximately 20 seconds and allows to consider the temperature field in the steady-state regime during the whole experiment: the longest time constant is due to thermal diffusion in air, only $\SI{50}{ms}$ on a conservative $\SI{1}{mm}$ scale.

\begin{table}[tb]
\center
\caption{\label{table::GeomCant} Cantilever geometries and resonant frequencies (in vacuum) for the three tested cantilevers. The geometrical dimensions were measured using a scanning electron microscope. The Reynolds number $\mathrm{Re}=\rho \omega b^2/4\eta$ associated with each resonance is computed using the properties of air at room temperature.}
\begin{tabular}{cccc}
  \hline
  \hline
   &  \hspace{4mm}Cant-A\hspace{4mm}   & \hspace{4mm}Cant-B\hspace{4mm} & \hspace{4mm}Cant-C\hspace{4mm} \\
  \hline
  $L$ ($\mu m$) & 505 & 354 & 508\\
  $b$ ($\mu m$) & 30 (14)\footnotemark[1] & 38 (31)\footnotemark[1] & 90 \\
  $h$ ($\mu m$)& 2.6 & 1.34 & 1.1 \\
  $L/b$ & 17 & 9 & 6 \\
  $b/h$ & 11 & 28 & 83 \\
     \hline
    $f_1$ (kHz)   & 15.2& 15.1 & 6.06\\
   $\mathrm{Re}$  & 1.4 & 2.3 & 5.1 \\
    $f_2$ (kHz)   & 95.6 & 95.2 & 37.9\\
     $\mathrm{Re}$  & 9 & 14 & 32\\
  $f_3$ (kHz)   & 266.4 & 266.4 & 105.9\\
       $\mathrm{Re}$  & 25 & 40 & 88\\
  $f_4$ (kHz)  & 521.6 & 521.7 & 207.2\\
       $\mathrm{Re}$  & 48 & 78 & 170\\
  $f_5$ (kHz)   & 860.6 & 860.9& 342.1\\
       $\mathrm{Re}$  & 80 & 130 & 280\\
      $f_6$ (kHz)   & - & - & 510.1\\      
      $\mathrm{Re}$  & - & - & 420\\
  \hline
  \hline
\end{tabular}
\footnotetext[1]{This cantilever has a trapezoidal cross section defined by two widths. Because the relevant length for the hydrodynamic problem is the maximum transverse dimension, the aspect ratios and the Reynolds number are computed using the maximum width.}
\end{table}

\subsection{Measurements in vacuum}

\begin{figure*}[htb]
 \centering
 \includegraphics[width=\textwidth]{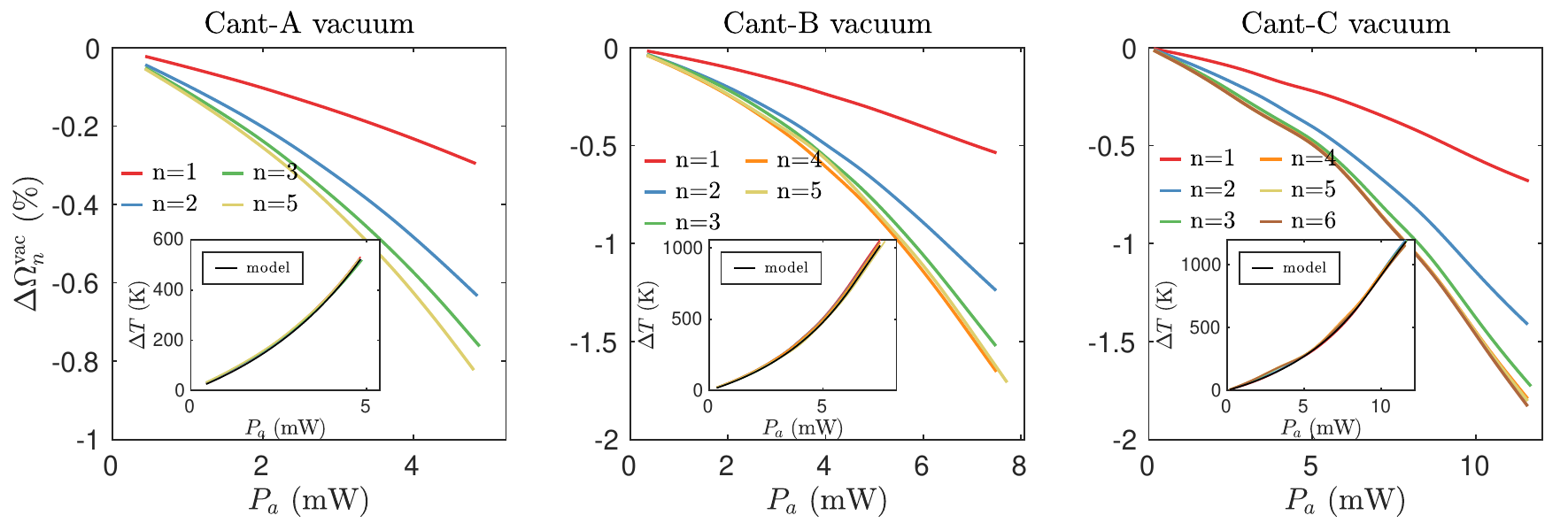}
 \includegraphics[width=\textwidth]{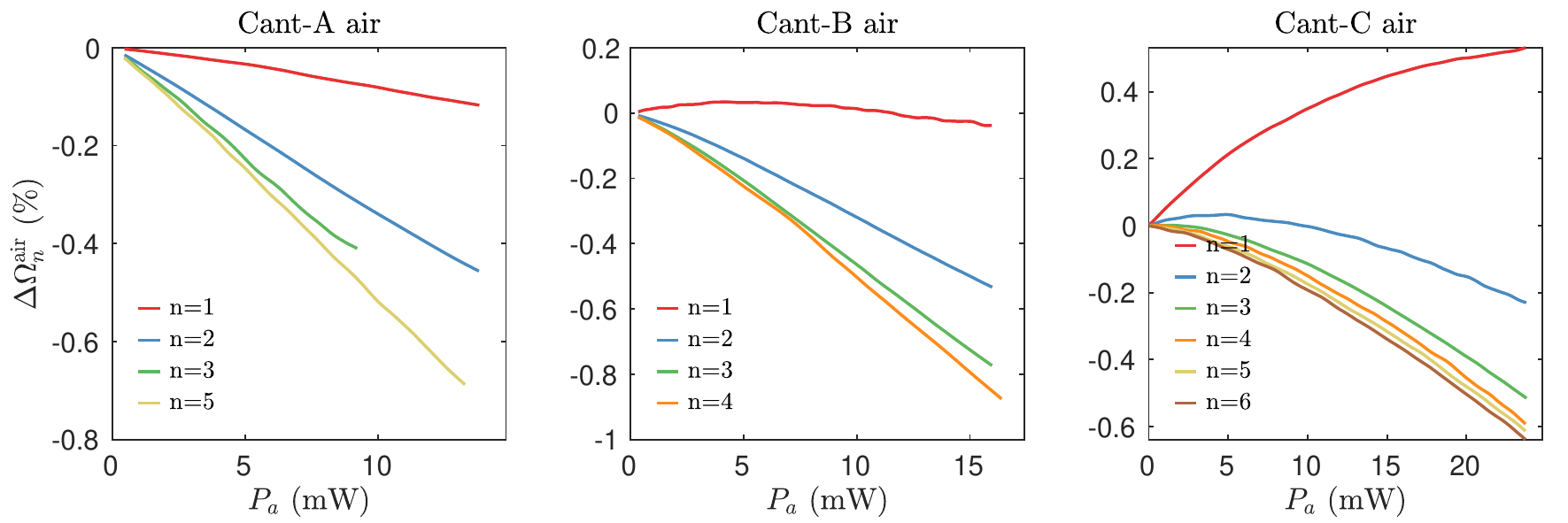}
   \caption{\label{Fig::ShiftVacAir} Measured relative frequency shifts in vacuum $\Delta \Omega_n^\mathrm{vac}$ (top) and in air at ambient pressure $\Delta \Omega_n^\mathrm{air}$(bottom) as a function of the absorbed power $P_a$. The cantilevers dimensions and resonant frequencies in vacuum at room temperature are listed in table~\ref{table::GeomCant}. While in vacuum the frequency shift for all modes is a monotonic function of absorbed power, in air the shift for lower modes can increase with $P_a$. Inset: temperatures rise $\Delta T$ deduced from the frequency shift for each mode measured in vacuum.}
\end{figure*}

In figure~\ref{Fig::ShiftVacAir} (top graphs), we display the relative frequency shift measured when the cantilever is placed in vacuum as a function of the absorbed power $P_a$. We were able to track the resonances up to the fifth mode for Cant-A and Cant-B and up to the sixth mode for Cant-C. As expected, the temperature increase induces a red-shift of its resonance frequencies. The shifts clearly depend on the mode being tracked. This mode dependence demonstrates that the induced cantilever temperature rise $\theta(x)$ is not uniform.

Knowing the temperature profile $\theta(x)=\Theta(x) \Delta T$ and the mechanical properties variation with temperature $E(T)$, $a_l(T)$, the relative frequency shift in vacuum $\Delta \Omega_n^\mathrm{vac}$ can be computed using equation Eq.~\eqref{eqShiftVac}. Conversely, as demonstrated in Ref.~\onlinecite{Pottier-2021-JAP}, knowing the normalized temperature profile $\Theta(x)$ and mechanical properties, the maximum temperature elevation $\Delta T$ can be extracted from the frequency shift $\Delta \Omega_n^\mathrm{vac}$ of any mode $n$
\begin{align}
	\Delta T=g_n^{-1} (\Delta \Omega_n^\mathrm{vac}).
\end{align}
The $g_n^{-1}$ functions are the inverse functions of $g_n$ which are defined according to Eq.~\eqref{eqShiftVac}. The normalized temperature profiles $\Theta(x)$ used to compute $g_n^{-1}$ can be obtained solving Eq.~\eqref{HeatDiffVac} using the silicon conductivity from Ref.~\onlinecite{glassbrenner_thermal_1964} and imposing the boundary conditions of Eqs.~\eqref{eq::CL_clamped}\footnote{The distance $l_\mathrm{th}$ used to compute the temperature profiles $\Theta(x)$ for Cant-A, Cant-B and Cant-C was respectively \SI{8}{\mu m}, \SI{8}{\mu m} and \SI{50}{\mu m}.} and \eqref{eq::CL_free}.
$\Delta T$ deduced for all tracked resonances is displayed in the insets of figure~\ref{Fig::ShiftVacAir}. All deduced temperatures for the three cantilevers are well superposed on the full range of absorbed power explored. They are also compared to the values predicted from Eq.~\eqref{HeatDiffVac}, with again an excellent agreement\footnote{The theoretical predictions displayed in the inset of figure~\ref{Fig::ShiftVacAir} was obtained using an effective thermal conductivity about 10\% lower than bulk silicon values. At micrometer scale, it has indeed been reported that the silicon conductivity is reduced due to phonon scattering at interfaces \cite{johnson_direct_2013,minnich_thermal_2011}.}.

As silicon is semi-transparent for visible light, the cantilever reflectivity results from the interferences between the multiple reflections within its thickness\cite{Pottier-2021-SciPost}. The refractive index of silicon is changing significantly with temperature, thus the interference state and the resulting reflection coefficient $R$ vary with $\Delta T$. Using the temperature deduced from the frequency shift, we plot the reflectivity of all three cantilevers in Fig.~\ref{fig::RvsT}: $R$ vary in an oscillatory fashion with a significant amplitude on the full temperature range. This signature is strongly dependent on the cantilever thickness, but once calibrated for a given sample, it can be used to retrieve the temperature at the laser spot position from the reflectivity measurement. As long as $\Delta T$ is varied in a continuous way in a single direction, the non-monotonicity of $R$ is not an issue in retrieving $\Delta T$. Thanks to this approach, we now have a thermometer for the cantilever tip not relying on the mechanical measurement.

\subsection{Measurements in air}

\begin{figure}[tb]
  \centering
  \includegraphics{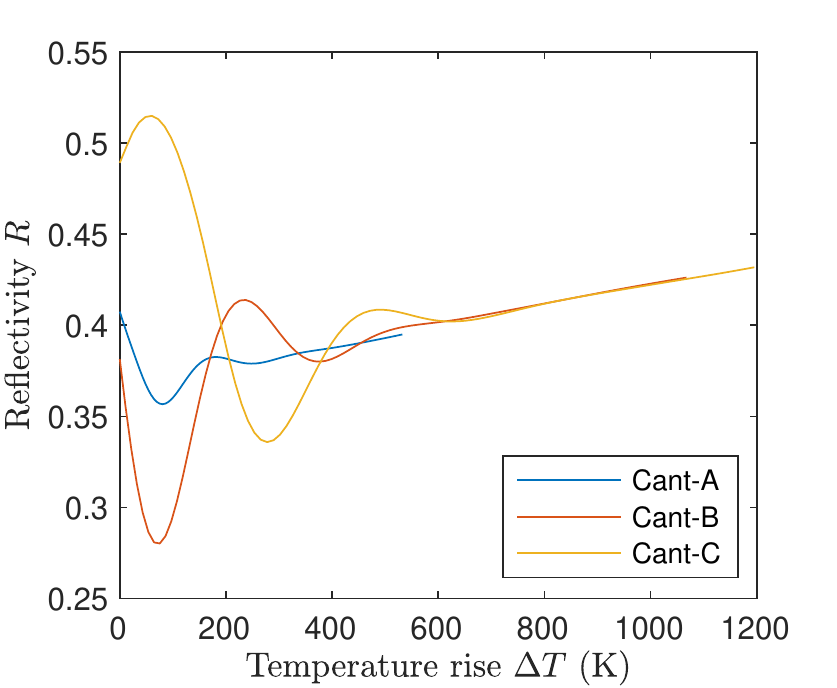} 
  \caption{\label{fig::RvsT} Reflectivity $R$ of the three cantilevers measured in vacuum as a function of the temperature rise $\Delta T$ at the laser spot position. $R$ results from multiple interferences in the cantilever thickness, and depends on temperature through the refractive index of silicon\cite{Pottier-2021-SciPost}.}
\end{figure}

\begin{figure}[t]
  \centering
  \includegraphics{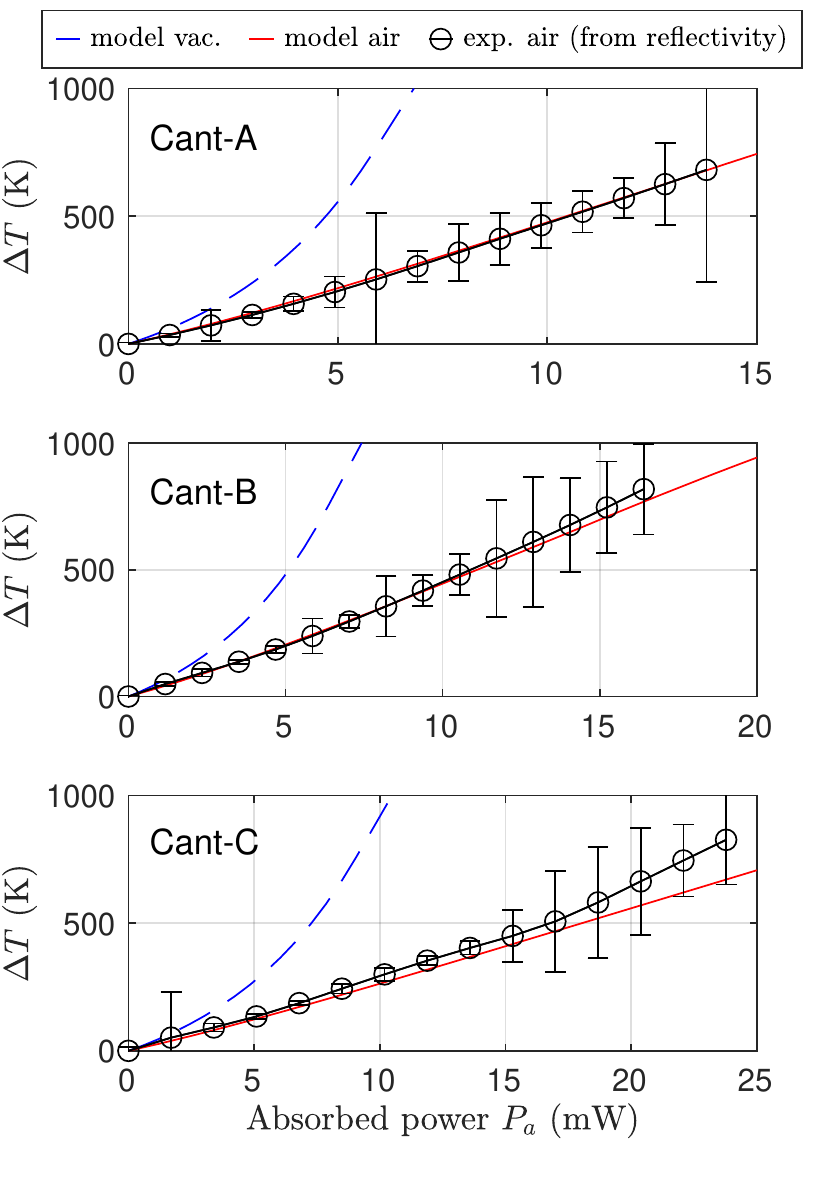} 
  \caption{\label{Fig::ExpComTempAirVac} Cantilever temperature rise measured in air from the reflectivity (symbols) and predicted in vacuum by Eq.~\eqref{HeatDiffVac} (blue line) and in air by Eq.~\eqref{eqHeatfluid} (red line). The temperature rise for the three cantilevers is well described by the presented model which takes into account the conduction through air. The theoretical blue lines correspond to the curve displayed in the insets of figure~\ref{Fig::ShiftVacAir}.}
\end{figure}

Similarly to the experiments in vacuum, the reflected intensities are measured as the cantilever is illuminated in air. Since variation in refractive index between air and vacuum is negligible, the reflectivity variations with temperature measured in air are identical to the ones recorded in vacuum. We can therefore use the vacuum measurement, reported in Fig.\ref{fig::RvsT} as a calibration, and determine the $\Delta T$ in the experiment performed in air from the measured reflectivity. The resulting temperature rise is plotted in Fig.~\ref{Fig::ExpComTempAirVac} as a function of the absorbed power $P_a$.

As expected, the presence of air, allowing an additional way to dissipate the absorbed heat, reduces the temperature rise (compared to the vacuum case): for the three cantilevers tested, the needed absorbed power to reach a temperature elevation is typically twice larger in air as the one needed in vacuum. The temperature rise measured in air from the reflectivity can be compared to the theoretical predictions made by our thermal model presented in section~\ref{section::thermalmodel}. The nice agreement obtained for all cantilevers validates our thermal model and confirms that the thermal dissipation in air is dominated by conduction.

\begin{figure}[htb]
  \centering
  \includegraphics[scale=.9]{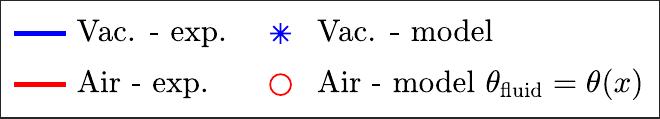}
  \centering
  \vspace{1mm}
 \includegraphics{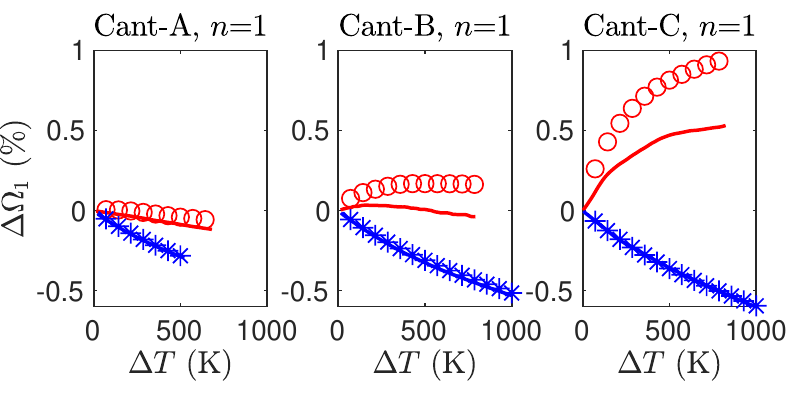} 
  \vspace{1mm}
  \includegraphics{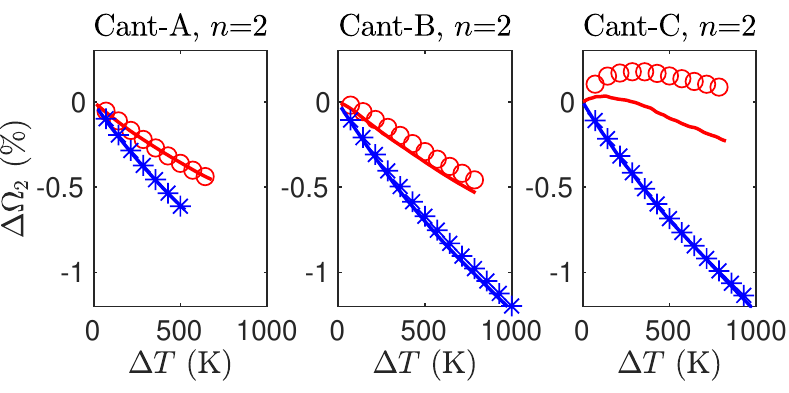} 
  \vspace{1mm}
 \includegraphics{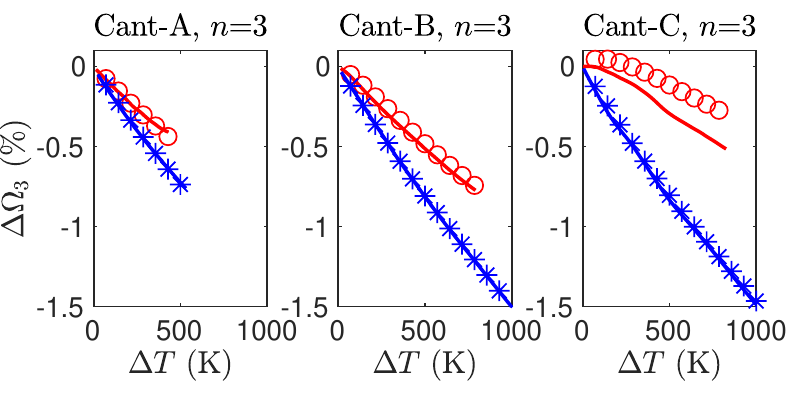}
  \caption{\label{fig::CompFreShift}Comparison of the relative frequency shift for the first three modes measured (lines) and theoretically predicted (symbols) for both vacuum (blue) and air (red). The effect of air decreases with the mode number (top to bottom) and increases with aspect ratio $b/h$ (left to right). The model predicting the frequency shift in air neglects the transverse temperature field in air, it only describes qualitatively the frequency shift.}
\end{figure}

In figure~\ref{Fig::ShiftVacAir} (bottom graphs), we display the frequency shift measured when the cantilever is placed in air. While in vacuum it is always decreasing with the absorbed power $P_a$, it increases in air for the first mode for cantilevers B and C (having larger aspect ratio $b/h$). As discussed in the previous paragraph, the temperature elevation induced for a given absorbed power is very different depending on the medium (air or vacuum) the cantilever is placed in. To compare more easily the frequency shift measured in both media, we display in figure~\ref{fig::CompFreShift} the frequency shift (for the first three modes) as a function of the temperature rise $\Delta T$ of the tip. As predicted in section~\ref{ModelFreqShift}, it is clear that the effect of the presence of air on the frequency shift increases with the aspect ratio $b/h$, and is much more pronounced for mode one.

We now compute the theoretical frequency shift in air using Eq.~\eqref{eqShiftAir}. In a first approach, we can suppose that the penetration depth of the air flow around the cantilever is very small, so that the fluid in the vicinity of the cantilever at the position $x$ can be supposed uniform at temperature $T_0+\theta(x)$. Under such hypothesis, the relative change of the dimensionless number $\zeta$ in Eq.~\eqref{eqShiftAir} can be estimated as 
\begin{align}\label{Azeta}
	A_\zeta(\theta) &= \frac{\zeta(\theta+T_0)}{\zeta(T_0)}-1.
\end{align}
The prediction of this approach is displayed in Fig.~\ref{fig::CompFreShift}: they describe qualitatively the experimental data although they overestimate systematically the effect of air.

From the analysis of thermal conduction in air of  section~\ref{section::thermalmodel}, we know that the fluid temperature actually decreases with the distance from the cantilever with a dominant length scale fixed by the width $b$. It is thus evident that the uniform model leading to Eq.~\eqref{Azeta} overestimates the effect of the heated air surrounding the cantilever. A more realistic estimation consists in determining the force exerted by the fluid on the cantilever using a nonuniform fluid temperature profile $T_\mathrm{fluid}(x,\vec{r})=T_0+\theta_\mathrm{fluid}(\vec{r})$
where $\vec{r}$ is the position in the transverse plane (y-z). In section~\ref{section::thermalmodel}, we analytically determine the temperature rise $\theta_\mathrm{fluid}$ for a circular cross section cantilever. For simplicity, we use this cylindrical geometry to evaluate the effect of the non-uniformity of the fluid temperature. At the first order, the fluid properties are expressed as
\begin{subequations}\label{Eq::PropAirVar}
\begin{align}
    \rho(x,r)&= \rho_0(1+ a_\rho \theta_\mathrm{fluid}(x,r)),  \\
    \eta(x,r)&= \eta_0(1+ a_\eta \theta_\mathrm{fluid}(x,r)),  
\end{align}
\end{subequations}
with $\theta_\mathrm{fluid}(x,r)$ given by Eq.~\ref{TempxsubL}. In the spirit of the 2-D Sader model\cite{sader_frequency_1998}, we then solve numerically the Navier-Stokes equations of a rigid cylinder oscillating at angular frequency $\omega$ in a nonuniform fluid. The computed force exerted by the fluid on the cylinder can be expressed by the same expression as for a uniform fluid Eq.~\eqref{EqForceHydro} in which $\rho$ and $\Gamma$ are computed at an effective temperature $T_\mathrm{fluid}^\mathrm{eff}$ given by 
\begin{align}
    T_\mathrm{fluid}^\mathrm{eff}=K(\omega) \theta(x)+T_0,
\end{align}
where the factor $K$ is a function of $\omega$, or equivalently the Reynolds number $\mathrm{Re}=\rho_0 \omega R^2/\eta_0$, and allows to takes into account the non-uniformity of the fluid temperature. Note that depending on which part of the force (real or imaginary) one wants to evaluate, there are two distinct factors to $K_r$ and $K_i$.

\begin{figure}[tb]
  \centering
 \includegraphics{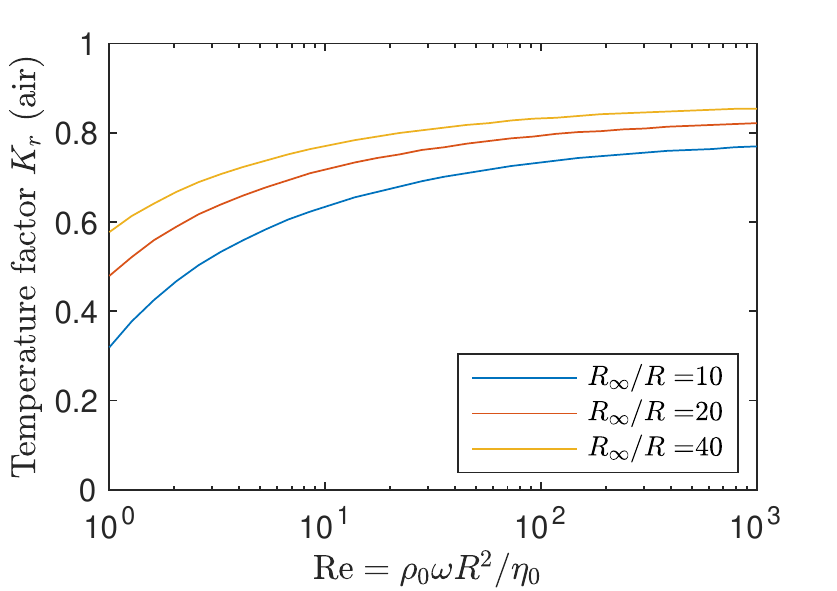} 
  \caption{\label{fig::FactorK} Coefficient $K_r$ to compute the equivalent uniform temperature of air when evaluating the real part of the hydrodynamic force exerted on a rigid cylinder oscillating perpendicularly to its axis. $K_r$ is obtained by solving the linearized Navier-Stokes equations with the temperature, density, and viscosity of air described by Eqs.~\eqref{TempxsubL} and \eqref{Eq::PropAirVar}. Owing to the larger volume involved in the flow in the viscous regime (lower $\mathrm{Re}$ values), the force is less sensitive to the temperature rise.}
\end{figure}

In Fig.~\ref{fig::FactorK}, we display the factor $K_r$ to use for the real part of the force using air properties ($a_\rho=\SI{-3.1e-3}{K^{-1}}$, $a_\eta=\SI{2.5e-3}{K^{-1}}$) for various factors $R_\infty/R$. The factor $K_r$ depends on the Reynolds number $\mathrm{Re}$. As the Reynolds number decreases, the fluid volume involved in the flow (owing to the viscous effect) increases. Thus, the force is more sensitive to the fluid properties in further regions where the temperature is lower, which translates into a lower effective temperature $T_\mathrm{fluid}^\mathrm{eff}$ (i.e. lower factor $K_r$). 

\begin{figure}[htb]
  \centering
  \includegraphics[scale=.83]{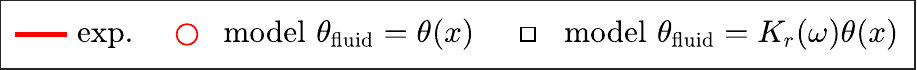} 
    \centering
   \vspace{3mm}
 \includegraphics{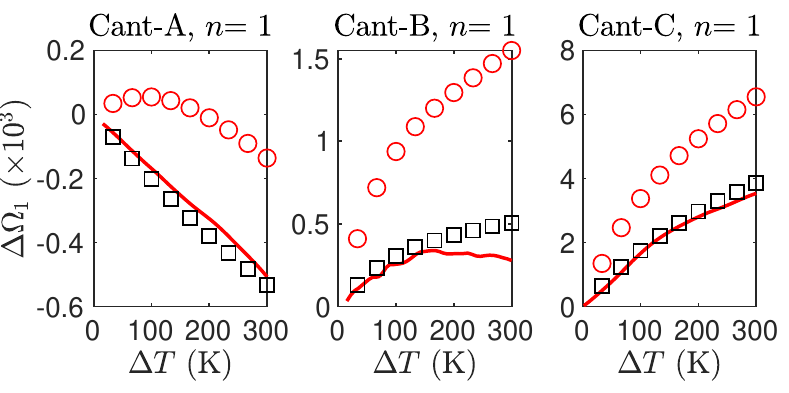} 
  \vspace{1mm}
  \includegraphics{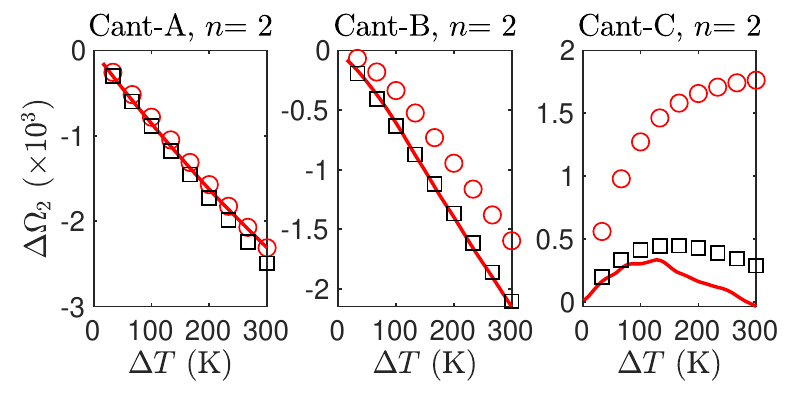} 
  \vspace{1mm}
 \includegraphics{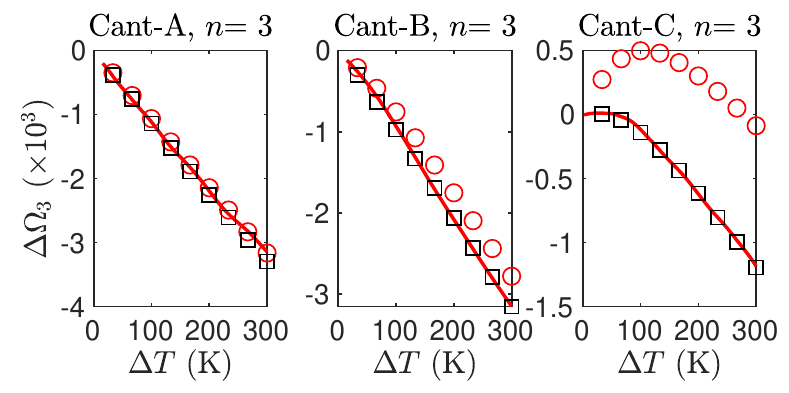}
  \caption{\label{fig::ShiftCompEquivTemp}Comparison of the frequency shift in air measured (line) and predicted from Eq.~\eqref{eqShiftAir} (symbols), neglecting the effect of the transverse temperature variation of air (Eq.~\ref{Azeta}) or not (Eq.~\ref{eq:AzetaKr}). This effect is significant (especially for the first mode) and has to be taken into account to describe the frequency shift.}
\end{figure}

The effect of the non-uniform air surrounding the cantilever can now be taken into account in the determination of the frequency shift using Eq.~\eqref{eqShiftAir}, using 
\begin{align} \label{eq:AzetaKr}
 	A_\zeta(\theta) &=\frac{\zeta(K_r(\omega)\theta+T_0)}{\zeta(T_0)}-1.
\end{align}
where $K_r$ is evaluated from the temperature field in air of Eq.\eqref{TempxsubL} with $R_\infty=R+x$. In contrast, the predictions presented in Fig.~\ref{fig::CompFreShift} assumed a fluid temperature equal to the one of the cantilever, thus $K_r=1$. In Fig.~\ref{fig::ShiftCompEquivTemp}, we display the frequency shift for the first three modes for a temperature elevation up to $\SI{300}{K}$. The comparison displayed on the full range of temperature explored is shown in appendix~\ref{AppendixCompFreqinAir}, Fig.~\ref{fig::Appendix}. All frequency shifts are better described taking into account the air temperature variation. The correction is the most important for the first mode for which the factor $K_r$ is the most different from 1: relative errors decrease from the 100\% range to a few percent for $\Delta T < \SI{150}{K}$. The remaining error, more pronounced for higher $b/h$ ratios, may be attributed to the oversimplification of the temperature field in air with the cylindrical cantilever boundary conditions. 

\section{Conclusion}

Let us summarise here the main conclusions of our work. First, we propose a model to describe the heat flow from a hot AFM cantilever to the surrounding medium and show that convective heat transfer is negligible at those scales. The heat transfer coefficient from the rigid surface to air is analytically computed in a cylindrical geometry, and extended to a rectangular cross-section cantilever. Corroborated by 3D numerical simulations and experiments on a wide temperature range, the model is quantitative in predicting the temperature profile as a function of heating power, with only slight deviations at high temperature for cantilevers with large width-to-thickness ratio $b/h$. As a rule of thumb for common AFM probe geometries, the heat flow in the environment is equal to the one through the cantilever: one needs to double the heating power to reach the same temperature when passing from vacuum to air.

We then study the interplay of the temperature field in air around the cantilever and the dynamics of the latter. To this aim, we track the effects on Sader's model induced by the reduced density and increased viscosity upon heating. We show that inertial effects decrease, tending to raise the resonance frequencies of the cantilever. The amplitude of this effect is of the same order of magnitude as the softening of the cantilever at high temperatures. The overall behavior is then geometry and mode dependant: for high width-to-thickness ratios $b/h$ and low mode number, the resonance frequency tends to increase, and conversely in the other directions. Those predictions match the experimental observations, with a quantitative agreement at moderate temperature rises. We demonstrate in the process that the temperature profile inside the fluid matters, since assuming uniform properties fails either way when considering the cantilever or the ambient temperature only.

We believe this work could have several applications when dealing with sensitive AFM measurements in the presence of temperature gradients. For example, the quantitative description of the heat flow in the environment can be useful to reach strong conclusions in thermal microscopy measurements, when performed in air. Our work for now only applies to the description of the probe far from a sample, and further work would be necessary to accommodate practical experimental configurations. Another important point is the influence of the temperature field when using dynamic modes, which are based on tracking resonance frequency: one should be aware that changes of the fluid properties have a noticeable (and traceable) influence. The fact that this effect is sensitive to boundary conditions, which change with respect to the present work in the vicinity of a sample, makes it harder to give general recipes, but at least some precautions should be taken in the result interpretations under this lighting.

Lastly, one could take advantage of the opposite effects of silicon softening and reduce gas inertial effects to tailor temperature sensitivity. If one wants for example to design fluid sensors, a high width-to-thickness ratio $b/h$ and low mode number are indicated to be sensitive to the gas temperature. On the contrary, using the right geometry for a target temperature range, one can compensate one effect with the other to reach temperature-insensitive probes as Cant. B for $\Delta T > \SI{200}{K}$ for mode 1, or Cant. C for $\SI{0}{K} <\Delta T < \SI{50}{K}$ for mode 3.

\subsection*{Acknowledgement}  
We thank Artyom Petrosyan and Jorge Pereda for enlightening technical and scientific discussions. Part of this research has been funded by the ANR project STATE (ANR-18-CE30-0013) of the Agence Nationale de la Recherche in France, and performed using the low noise chamber of the Optolyse platform, funded by the r\'egion Auvergne-Rh\^one-Alpes (CPER2016).

\subsection*{Data availability}
The data that support the findings of this study are openly available in Zenodo at \url{https://doi.org/10.5281/zenodo.5346796}~\cite{Pottier-2021-air-Dataset-JAP}

\subsection*{Conflict of interest}
The authors have no conflicts to disclose.

\bibliography{ArticleShiftAir}

\begin{thebibliography}{41}
\bibitem [{Not()}]{NoteMissingQ}%
  \BibitemOpen
  \href@noop {} {}\bibinfo {note} {{O}wing to the proximity with a torsional
  resonance, we were not able to measure the quality factor of the third
  (flexural) resonance for Cant-B ($b/h=28$).}\BibitemShut {Stop}%
\end{thebibliography}

\appendix

\clearpage
\onecolumngrid

\section{Comparison of the frequency shift in air}  \label{AppendixCompFreqinAir} 

\begin{figure*}[htb]
    \includegraphics{legend_zoom.pdf}
    
    \hspace{3mm}
    
    \includegraphics[height=199mm]{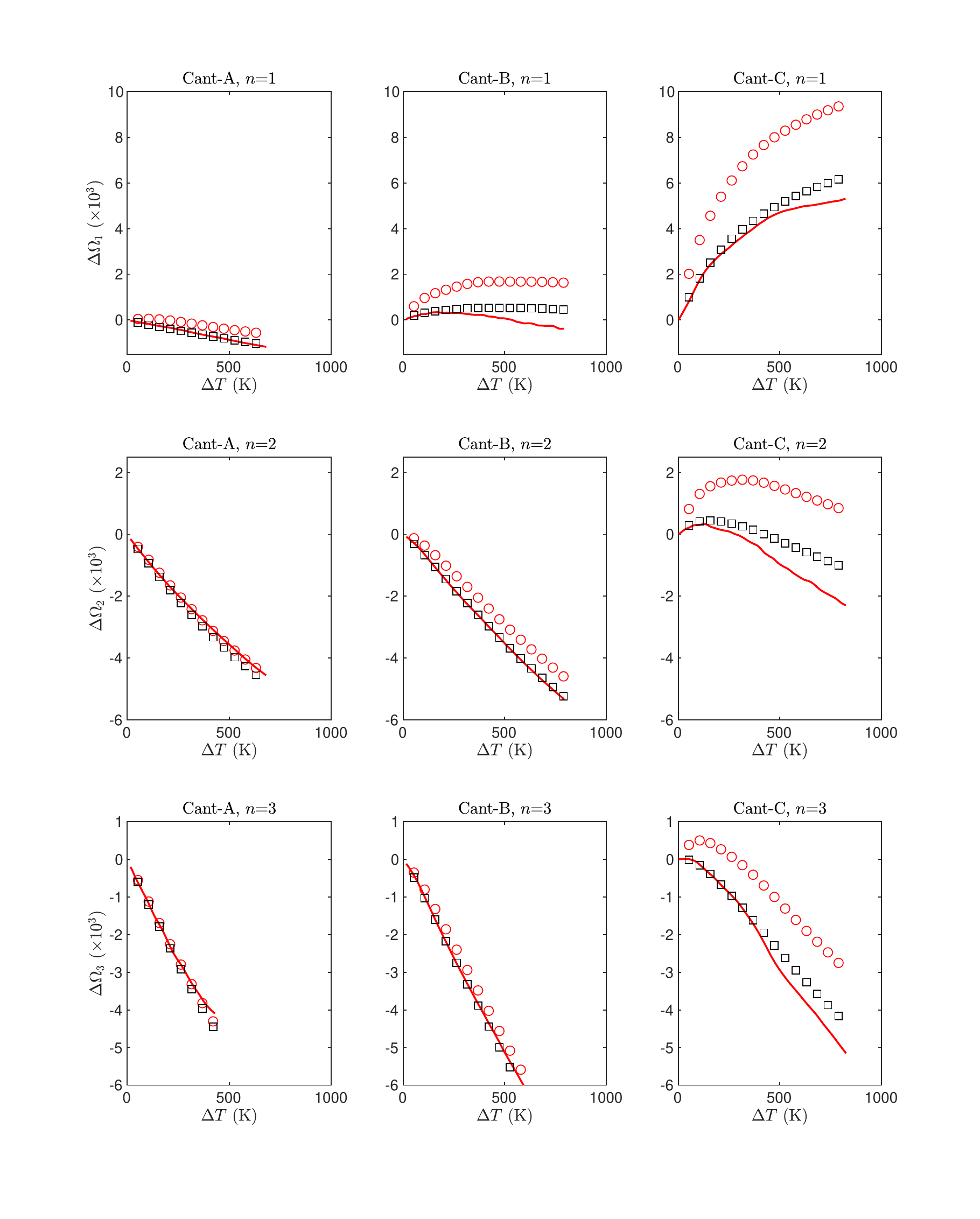}
    \caption{\label{fig::Appendix}Comparison of the frequency shift in air measured (line) and predicted from Eq.~\eqref{eqShiftAir} (symbols), neglecting the effect of the transverse temperature variation of air (Eq.~\ref{Azeta}) or not (Eq.~\ref{eq:AzetaKr}).}
\end{figure*}

\end{document}